\documentclass{aastex631}
\usepackage{newtxtext,newtxmath}
\shorttitle{Scattering blanketing effect of Earth's proto-atmosphere}
\shortauthors{Yoshida et al.}
\graphicspath{{./}{figures/}}
\begin{document}

\title{Scattering blanketing effect of Earth's proto-atmosphere: enhanced suppression of planetary radiation and magma ocean cooling}

\author{Tatsuya Yoshida}
\affiliation{Faculty of Science, Tohoku University, Sendai, Miyagi 980-8578, Japan}

\author{Kirara Arima}
\affiliation{Faculty of Science, Hokkaido University, Sapporo, Hokkaido 060-0810, Japan}

\author{Takeshi Kuroda}
\affiliation{Faculty of Science, Tohoku University, Sendai, Miyagi 980-8578, Japan}
\affiliation{Division for the Establishment of Frontier Sciences of Organization for Advanced Studies, Tohoku University, Sendai, Miyagi 980-8577, Japan}

\author{Naoki Terada}
\affiliation{Faculty of Science, Tohoku University, Sendai, Miyagi 980-8578, Japan}

\author{Kiyoshi Kuramoto}
\affiliation{Faculty of Science, Hokkaido University, Sapporo, Hokkaido 060-0810, Japan}



\begin{abstract}
	The thermal evolution of magma oceans formed by giant impacts is strongly influenced by a proto-atmosphere through its blanketing effect, which suppresses outgoing planetary radiation. While both radiative absorption and Rayleigh scattering by atmospheric species can contribute to this effect, the role of the scattering in suppressing thermal radiation from magma oceans remains unclear. In this study, we developed a 1-D radiative transfer model for planetary and solar radiation in a proto-atmosphere composed of H$_2$O and H$_2$, and a coupled thermal evolution model of a planetary interior and proto-atmosphere, to investigate the scattering blanketing effect on planetary radiation and magma ocean cooling. Our results show that Rayleigh scattering significantly reduces outgoing planetary radiation at wavelengths below $\sim 1\,\mathrm{\micron}$, particularly in hot, thick atmospheres where scattering is highly effective. Consequently, the planetary outgoing radiation flux decreases by up to about one to two orders of magnitude, and the magma ocean lifetime is prolonged by up to about three times due to the scattering blanketing effect when the total amounts of H$_2$O and H$_2$ are equivalent to or greater than the present-day terrestrial seawater. These findings suggest that the prolonged magma ocean phase facilitated efficient differentiation between compatible and incompatible elements, even in the lower mantle. Furthermore, they imply that sustained magma oceans likely persisted throughout much of the giant impact phase, supporting a magma ocean origin of the Moon consistent with its observed chemical characteristics.
\end{abstract}



\section{Introduction} \label{sec:intro}	
	Planet formation theories suggest that Earth-sized planets likely underwent a globally molten state due to giant impacts during the late stages of accretion \citep[e.g.,][]{Canup2004a, Zahnle2007, Solomatov2007,Elkins2012}. This magma ocean would have cooled primarily through thermal radiation into space. The cooling process is known to be strongly influenced by the properties of proto-atmospheres, particularly through their blanketing effects. The role of steam atmospheres in this context has been extensively studied \citep[][]{Abe1985, Abe1986, Abe1988, Matsui1986a, Matsui1986b, Zahnle1988, Elkins2008, Elkins2012, Lebrun2013, Hamano2013, Hamano2015, Massol2016, Massol2023, Schaefer2016, Salvador2017, Marcq2017, Ikoma2018, Bower2019, Katyal2019, Nikolaou2019, Pluriel2019, Barth2021, Lichtenberg2021}. These atmospheres can drastically lower the outgoing planetary radiation flux by several orders of magnitude compared to blackbody radiation, due to the infrared-absorbing properties of H$_2$O \citep[e.g.,][]{Kasting1988, Nakajima1992, Goldblatt2013}. Consequently, the presence of thick steam atmospheres with masses comparable to or exceeding Earth's current ocean mass can extend the lifetime of the magma ocean to several million years at Earth’s orbital distance \citep[e.g.,][]{Lebrun2013, Hamano2013}.
	
	Recent studies have examined magma ocean evolution across a wide range of atmospheric compositions by considering diverse oxygen fugacities in magma ocean-atmosphere systems \citep{Katyal2020, Lichtenberg2021, Bower2022, Maurice2024, Nicholls2024, Nicholls2025}. \citet{Lichtenberg2021} investigated the blanketing effects of individual atmospheric species of H$_2$O, H$_2$, CO$_2$, CH$_4$, CO, O$_2$, and N$_2$ by modeling single-component atmospheres. Their findings indicate that H$_2$ exhibits particularly strong blanketing effects due to collision-induced absorption at high-pressure conditions: the solidification of the magma ocean beneath an H$_2$ atmosphere with a mass of the present-day terrestrial seawater is not completed within 100 million years even at the Earth's orbit.
	
	In addition to radiative absorption, scattering of planetary radiation can also contribute to the blanketing effect. Rayleigh scattering becomes significant at shorter wavelengths; for instance, the Rayleigh scattering cross-section of H$_2$O exceeds its absorption cross-section at wavelengths below $\sim 0.5\,\mathrm{\micron}$ \citep[e.g.,][]{Pierrehumbert2010, Rothman2013}. As suggested by the blackbody spectrum, the contribution of thermal radiation at such short wavelengths, influenced by Rayleigh scattering, increases with the surface temperature. Consequently, Rayleigh scattering by atmospheric species is expected to reduce outgoing planetary radiation from a hot magma ocean, potentially extending the magma ocean’s lifetime. Although some previous studies on the thermal evolution with steam atmospheres considered the Rayleigh scattering of planetary radiation \citep[e.g.,][]{Abe1988, Zahnle1988, Goldblatt2013, Hamano2015}, the specific impact on thermal radiation from magma oceans remains unclear.

	In this study, we developed a 1-D radiative transfer model for planetary and solar radiation in a proto-atmosphere composed of H$_2$O and H$_2$, incorporating both radiative absorption and scattering processes, to investigate the scattering blanketing effect on thermal planetary radiation. Additionally, to estimate the thermal evolution of a magma ocean and proto-atmosphere, we coupled the radiative transfer model with a planetary interior model that accounts for thermal structure and volatile partitioning. The remainder of this paper is organized as follows: in Section 2, we describe our 1-D radiative transfer model and the coupled thermal evolution model of the magma ocean and atmosphere. Section 3.1 presents the numerical results for planetary and solar radiation transfer in proto-atmospheres composed of H$_2$O and H$_2$, with a focus on the scattering blanketing effect. Section 3.2 shows the thermal co-evolution of an H$_2$O-H$_2$ proto-atmosphere and magma ocean. Section 4.1 examines the effects of uncertainties in atmospheric structure assumptions. Section 4.2 discusses the chemical evolution of proto-atmosphere and magma ocean. Section 4.3 describes the effects of our results on chemical differentiation in a magma ocean. Section 4.4 discusses the implications of our results for the Moon’s origin.

\section{Model description} \label{sec:model}
\subsection{Atmospheric structure}
We suppose that atmospheres are composed of H$_{2}$O and H$_{2}$, which are expected to be major radiative absorption sources in proto-atmospheres \citep{Lichtenberg2021}. The atmosphere in hydrostatic equilibrium is vertically divided into 500 layers from the ground to the top of the atmosphere, which are equally spaced on a logarithmic scale of pressure. The pressure at the top of the atmosphere is fixed at 1 Pa. We treat H$_{2}$ as an ideal gas and H$_{2}$O as a non-ideal gas following the Peng-Robinson equation of state \citep{Peng1976}, in a similar manner to \citet{Abe1988} and \citet{Hamano2015}. The Peng-Robinson equation of state is described as follows:
	\begin{equation}
		P=\frac{RT}{V-b}-\frac{a(T)}{V(V+b)+b(V-b)},
	\end{equation}
	where $P$ is the pressure, $R$ is the gas constant, $T$ is the temperature, and $V$ is the molar volume \citep{Peng1976}. Here $a(T)$ and $b$ are defined by
	\begin{equation}
		a(T)=0.45724\frac{R^{2}T_{c}^{2}}{P_{c}}\left[1+\kappa \left(1-\sqrt{\frac{T}{T_{c}}}\right)\right]^{2},
	\end{equation}
	\begin{equation}
		b=0.07780\frac{RT_{c}}{P_{c}},
	\end{equation}
	\begin{equation}
		\kappa = 0.37464 + 1.54226\omega_{a} - 0.26992\omega_{a}^{2},
	\end{equation}
	where $T_{c}$ and $P_{c}$ are the temperature and pressure at the critical point, respectively, and $\omega_{a}$ is the acentric factor. Here $P_{c}=2.204\times 10^{7}\,\mathrm{Pa}$, $T_{c}=647.3\,\mathrm{K}$, and $\omega_{a}=0.344$ for H$_{2}$O \citep[e.g.,][]{Chase1998}.

	The surface temperature is varied from 200 K to 4000 K during the course of the radiative transfer calculations. The temperature profile in the lower part of the atmosphere is supposed to follow an adiabatic lapse rate, and that in the upper part is given by the skin temperature, which is the asymptotic temperature at high altitudes of an upper atmosphere that is optically thin in the thermal-IR and transparent to shortwave radiation. The lapse rate is assumed to be a dry adiabat for atmospheric layers in which the temperature is above the critical temperature of H$_{2}$O or the relative humidity is less than 1. The dry adiabatic lapse rate is given by
	\begin{equation}
		\left(\frac{\partial T}{\partial P}\right)_{\rm dry}=\frac{\rho_{v}T}{\rho_{v}C_{p,v}+\rho_{n}C_{p,n}}\left(\frac{\partial (1/\rho_{v})}{\partial T}\right)_{P},
	\end{equation}
	where $\rho_{v}$ and $\rho_{n}$ are the densities of the condensible gas and non-condensible gas, respectively, $C_{p,v}$ and $C_{p,n}$ are the constant-pressure specific heat of the condensible gas and non-condensible gas, respectively \citep{Kasting1988, Marcq2012, Marcq2017}. For the layers where the relative humidity is unity, the lapse rate is given by a pseudo-moist adiabat as given by
	\begin{equation}
		\left(\frac{\partial T}{\partial P}\right)_{\rm moist}=\left[\frac{dP_{s}}{dT}+\frac{\rho_{n}R}{M_{n}}\left(1+\frac{d \mathrm{ln}\rho_{v}}{d \mathrm{ln}T}-\frac{d \mathrm{ln}(\rho_{v}/\rho_{n})}{d \mathrm{ln}T}\right)\right]^{-1},
	\end{equation}
	where $P_{s}$ is the saturation vapor pressure of H$_{2}$O and $M_{n}$ is the molecular weight of the non-condensible gas \citep{Kasting1988, Marcq2012, Wordsworth2013, Marcq2017}. Here,
	\begin{equation}
		\frac{d \mathrm{ln}(\rho_{v}/\rho_{n})}{d \mathrm{ln}T}=\frac{1}{(\rho_{v}/\rho_{n})(s_{v}-s_{c})+R/M_{n}}\left[\frac{R}{M_{n}}\frac{d \mathrm{ln}\rho_{v}}{d \mathrm{ln}T}-C_{v,n}-\frac{\rho_{v}}{\rho_{n}}\left(\frac{d s_{v}}{d \mathrm{ln}T}\right)\right],
	\end{equation}
	where $s_{v}$ and $s_{c}$ are the specific entropies of the vapor phase and the condensed phase, respectively \citep{Kasting1988}. When the temperature reaches the skin temperature, the temperature above is assumed to be constant at the skin temperature as given by
	\begin{equation}
		T_{\rm skin}=\left(\frac{F_{\rm OLR}}{2\sigma}\right)^{1/4},
	\end{equation}
	where $\sigma$ is the Stefan-Boltzmann constant and $F_{\rm OLR}$ is the outgoing longwave radiation flux. The method to obtain $F_{\rm OLR}$ is described in the next subsection. We conducted iterative procedures to calculate $T_{\rm skin}$ and $F_{\rm OLR}$ self-consistently. 
	
	We refer to the temperature-dependent specific heat, entropy, and latent heat data provided in the NIST Chemistry WebBook \citep{Chase1998}. The saturation vapor pressure of H$_{2}$O is given by Osborne-Mayers and Washburn formula in the same manner as \citet{Abe1988}. At $273.15\,\mathrm{K} \leq T \leq 647.26\,\mathrm{K}$, 
	\begin{equation}
		\mathrm{ln}P_{s}(T)=\alpha - \frac{\beta}{t}+\frac{\gamma x}{t}[\mathrm{exp}(\delta x^{2})-1]-\epsilon \mathrm{exp}(-\eta y),
	\end{equation}
	where $t=T+0.01$, $x=t^{2}-293700$, $y=(647.26-T)^{5/4}$, $\alpha=24.021415$, $\beta=4616.9134$, $\gamma = 3.1934553\times 10^{-4}$, $\delta=2.7550431\times 10^{-11}$, $\epsilon=1.0246503\times 10^{-2}$, and $\eta=1.3158813\times 10^{-2}$ \citep{Eisenberg1969, Abe1988}. At $T \leq 273.15\,\mathrm{K}$,
	\begin{equation}
		\mathrm{ln}P_{s}(T)=\alpha'-\frac{\beta'}{t}+\gamma'\mathrm{log}t-\delta't+\epsilon't^{2},
	\end{equation}
	where $t=T-0.05$, $\alpha'=-10.666189$, $\beta'=5631.1206$, $\gamma' =18.953038$, $\delta'=3.8614490\times 10^{-2}$, and $\epsilon'=2.7749374\times 10^{-5}$ \citep{Abe1988}. 
	
\subsection{Radiative transfer calculations}
	We solve the radiative transfer equation in a plane-parallel atmosphere on the two-stream approximation \citep{Toon1989} for the planetary and stellar radiation separately:
	\begin{equation}
		\frac{\partial F_{\nu}^{+}}{\partial \tau_{\nu}}=\gamma_{1}F_{\nu}^{+}-\gamma_{2}F_{\nu}^{-}-S_{\nu}^{+},
	\end{equation}
	\begin{equation}
		\frac{\partial F_{\nu}^{-}}{\partial \tau_{\nu}}=\gamma_{2}F_{\nu}^{+}-\gamma_{1}F_{\nu}^{-}+S_{\nu}^{-},
	\end{equation}
	where $\nu$ is the wavenumber, $\tau_{\nu}$ is the optical depth at $\nu$, $F_{\nu}^{+}$ and $F_{\nu}^{-}$ are the upward and downward radiative fluxes at $\nu$, and $\gamma_{i}$ is the coefficient which depends upon the particular form of the two-stream equation, respectively. For the planetary radiation,
	\begin{equation}
		S_{\nu}^{+}=S_{\nu}^{-}=2\pi (1-\omega_{0})B_{\nu}(T),
	\end{equation}
	where $\omega_{0}$ is the single scattering albedo and $B_{\nu}(T)$ is the Planck function. For the solar flux,
	\begin{equation}
		S_{\nu}^{+}=\gamma_{3}F^{s}_{\nu}\omega_{0}\mathrm{exp}\left(-\frac{\tau_{\nu}}{\mu_{0}}\right),
	\end{equation}
	\begin{equation}
		S_{\nu}^{-}=(1-\gamma_{3})F^{s}_{\nu}\omega_{0}\mathrm{exp}\left(-\frac{\tau_{\nu}}{\mu_{0}}\right),
	\end{equation}
	where $F^{s}_{\nu}$ is the incoming solar flux at $\nu$, $\tau_{\nu}$ is the optical depth from the top of the atmosphere at $\nu$, and $\mu_{0}$ is the cosine of the mean stellar zenith angle ($57.3^{\circ}$) \citep[e.g.,][]{Hu2012}. Following the recommendation by \citet{Toon1989}, we use the hemispheric mean method for planetary radiation and the quadrature method for solar radiation. Here $\gamma_{1}=2-\omega_{0}$ and $\gamma_{2}=\omega_{0}$ in the hemispheric mean method, and $\gamma_{1}=\sqrt{3}(2-\omega_{0})/2$, $\gamma_{2}=\sqrt{3}\omega_{0}/2$, and $\gamma_{3}=1/2$ in the quadrature method \citep{Toon1989}.

	As radiative absorption sources, we consider line and continuum absorption of H$_{2}$O and collision-induced absorption of H$_{2}$-H$_{2}$. We use the H$_{2}$O line data in HITEMP 2010 \citep{Rothman2010}. The line profile of H$_{2}$O is assumed to follow a Voigt profile, truncated at 25 cm$^{-1}$ from the line center, to combine with continuum absorption represented by the MT\_CKD 4.2 model \citep{Mlawer2023} without double-counting. Here we use EXOCROSS \citep{Yurchenko2018} for the line profile calculations. We use the H$_{2}$-H$_{2}$ collision-induced absorption data provided by \citet{Borysow2001} and \citet{Borysow2002}. We also consider Rayleigh scattering by H$_{2}$O and H$_{2}$. The scattering cross-section is given by
	\begin{equation}
		\sigma^{s}_{i}(\lambda)=\frac{128\pi^{5}(\alpha_{i}^{p})^{2}}{3\lambda^{4}},
	\end{equation}
	where $\lambda$ is the wavelength, and $\sigma^{s}_{i}$ and $\alpha^{p}_{i}$ are the scattering cross-section and the polarizability constant of species $i$, respectively \citep[e.g.,][]{Pierrehumbert2010}. Here $\alpha^{p}_{\rm H_{2}O}=1.50$ \AA$^3$ and $\alpha^{p}_{\rm H_{2}}=0.787$ \AA$^3$, respectively \citep{Johnson1998}. Considering the absorption and scattering by the atmosphere, the optical depth $\tau_{\nu}$ is calculated by
	\begin{equation}
		\frac{d\tau_{\nu}}{dP}=\frac{1}{\rho g}\left[ \sum_{i} (\sigma^{a}_{i}(\nu)+\sigma^{s}_{i}(\nu))n_{i}+k_{\rm H_{2}-H_{2}}(\nu)n_{\rm H_{2}}^{2}\right],
	\end{equation}
	where $\rho$ is the atmospheric mass density, $g$ is the gravitational acceleration, $n_{i}$ is the number density of species $i$, $\sigma^{a}_{i}(\nu)$ and $\sigma^{s}_{i}(\nu)$ are the absorption and scattering cross-sections at $\nu$ of species $i$, and $k_{\rm H_{2}-H_{2}}(\nu)$ is the absorption coefficient of the H$_{2}$-H$_{2}$ collision-induced absorption at $\nu$, respectively.
	
	By calculating the scattered upward flux at the top of the atmosphere, we estimate the planetary albedo defined as follows:
	\begin{equation}
		A_{p}=\frac{F^{+}_{s,\mathrm{top}}}{\mu_{0}S_{\odot}},
	\end{equation}
	where $A_{p}$ is the planetary albedo, $F^{+}_{s,\mathrm{top}}$ is the scattered upward flux at the top of the atmosphere, and $S_{\odot}$ is the solar constant. Here we adopt the solar constant 0.7 times the present-day value considering the faint young Sun \citep{Gough1981}. The surface albedo is assumed to be 0.2 \citep[e.g.,][]{Kasting1988}. Note that the radiative effects of clouds are not included in this model, which may cause the estimated planetary albedo to deviate from the true 3-D average planetary albedo (see Section 4.1.3). The net absorbed solar flux is given by
	\begin{equation}
		F_{\rm Sun}=\frac{1}{4}(1-A_{p})S_{\odot}.
	\end{equation}
		
	We evaluate the integral of the radiative flux over a wavenumber range from 1 cm$^{-1}$ to 30000 cm$^{-1}$ with a resolution of 1 cm$^{-1}$ by applying a line-by-line method following \citet{Schaefer2016} and \citet{Wordsworth2017}. We tabulate the absorption cross-section on temperature and pressure grids. Here the temperature grids are [100, 150, 200, 250, 300, 350, 400, 450, 600, 800, 1000, 1250, 1500, 1750, 2000, 2250, 2500, 2750, 3000, 3250, 3500, 3750, 4000] K, and the pressure grids are [$10^{-5}$, $10^{-4}$, $5\times 10^{-4}$, $10^{-3}$, $5\times 10^{-3}$, $10^{-2}$, $5\times 10^{-2}$, $10^{-1}$, $5\times 10^{-1}$, 1, 5, 10, 50, $10^{2}$, $5\times 10^{2}$, $10^{3}$, $5\times 10^{3}$] bar. 
	
\subsection{Interior structure and energy conservation}
	In Section 3.2, we estimate the coupled thermal evolution of the atmosphere and the interior by combining the radiative flux depending on the surface temperature and atmospheric amount and composition. We consider a spherically symmetric hydrostatic mantle where the lower boundary is the core-mantle boundary with a depth of 2900 km and the upper boundary is the planetary surface. The mantle is initially assumed to be fully molten with a potential temperature of 4000 K. The magma ocean is convecting vigorously, so the temperature distribution is supposed to be adiabatic following \citet{Lebrun2013}. As it cools, the adiabatic temperature profile intersects with the liquidus and solidus temperature profiles from the bottom. We use the solidus and liquidus curves of KLB-1 peridotite obtained from experimental data \citep{Hirschmann2000, Herzberg2000, Fiquet2010, Zhang1994}, whose fitting formula is derived by \citet{Nikolaou2019}. The melt fraction at any depth is given by
	\begin{equation}
		\phi = \frac{T-T_{\rm sol}}{T_{\rm liq}-T_{\rm sol}},
	\end{equation}
	where $\phi$ is the melt fraction, $T$ is the temperature of the mantle at a given depth, $T_{\rm sol}$ is the solidus temperature, and $T_{\rm liq}$ is the liquidus temperature, respectively \citep[e.g.,][]{Abe1997, Solomatov2007, Lebrun2013, Nikolaou2019}. The adiabatic temperature gradient is given by
	\begin{equation}
		\left(\frac{\partial T}{\partial P}\right)_{s} = \begin{cases}
			\frac{\alpha_{T} T}{\rho_{m}C_{p,M}},\hspace{10pt}\mathrm{for}\,\phi=1 \\
			\frac{\alpha_{T} T}{\rho_{s}C_{p,M}},\hspace{10pt}\mathrm{for}\,\phi=0 \\
			\frac{\alpha_{T} T/\rho_{M}-\Delta H (\partial \phi/\partial P)_{T}}{C_{p,M}-\Delta H (\partial \phi/\partial T)_{P}},\hspace{10pt}\mathrm{for}\,0<\phi<1 \\
  		\end{cases}
	\end{equation}
	where $\rho_{m}$ is the melt density ($=4.0\times 10^{3}\,\mathrm{kg/m^{3}}$), $\rho_{s}$ is the solid density ($=4.2\times 10^{3}\,\mathrm{kg/m^{3}}$), $\rho_{M}$ is the mean density of the melt and solid, $C_{p,M}$ is the specific heat of the mantle ($=1.0\times 10^{3}\,\mathrm{J/kg/K}$), $\Delta H$ is the difference of specific enthalpy between the melt and solid phases ($=4\times 10^{5}\,\mathrm{J/kg}$), and $\alpha_{T}$ is the thermal expansivity given by
	\begin{equation}
		\alpha_{T}=\alpha_{T,0}\left(\frac{PK'}{K_{0}}+1\right)^{-(m-1+K')/K'},
	\end{equation}
	where $\alpha_{T,0}=3\times 10^{-5}\,\mathrm{K^{-1}}$, $K_{0}=200\,\mathrm{GPa}$, $K'=4$, and $m=0$, respectively \citep{Abe1997, Lebrun2013}.
	
	The coupled thermal evolution is obtained by the energy conservation equation:
	\begin{equation}
		r^{2} \rho_{M} \left[C_{p,M}+\Delta H \left(\frac{\partial \phi}{\partial T}\right)_{P}\right]\frac{\partial T}{\partial t}=\frac{\partial}{\partial r}[r^{2}(F_{\rm Sun}-F_{\rm OLR})],
	\end{equation}
	where $r$ is the radial distance from the planetary center and $F_{\rm OLR}$ is the outgoing planetary radiation flux, respectively \citep{Lebrun2013}. Here the cooling from the magma ocean is supposed to be along with the net energy balance of the incoming and outgoing radiation at the top of the atmosphere \citep{Hamano2013, Katyal2019}. 
	
\subsection{Partitioning of volatiles between the atmosphere and the interior}
	In Section 3.2, the partitioning of volatiles among the solid mantle, magma ocean, and atmosphere is considered as follows:
	\begin{equation}
		k_{i}X_{i}M_{\rm solid}+X_{i}M_{\rm liquid}+\frac{4\pi R_{p}^{2}}{g}\frac{\mu_{i}}{\bar{\mu}}P_{i}=M^{0}_{i},
	\end{equation}
	where $R_{p}$ is the planetary radius, $X_{i}$, $P_{i}$, $M^{0}_{i}$, $\mu_{i}$, and $k_{i}$ are the mass fraction in the magma ocean, atmospheric partial pressure, initial total amount, molecular weight, and distribution coefficient between the solid and liquid phases of species $i$, respectively, $\bar{\mu}$ is the mean molecular weight of the atmosphere, $M_{\rm solid}$ is the mass of the solid mantle, and $M_{\rm liquid}$ is the mass of the magma ocean \citep{Lebrun2013, Bower2019, Nikolaou2019}. We assume that $k_{\rm H_{2}O}=k_{\rm H_{2}}=0.01$ for simplicity by referring to the values for H$_2$O used in previous studies \citep[e.g.,][]{Lebrun2013}. We use a solubility law to relate the atmospheric partial pressure and the mass fraction in the magma ocean as follows:
	\begin{equation}
		P_{i} = \left(\frac{X_{i}}{\alpha_{i}}\right)^{\beta_{i}},
	\end{equation}
	where $\alpha_{i}$ and $\beta_{i}$ are the Henrian fit coefficients of species $i$. Here $\alpha_{\rm H_{2}O}=6.8\times 10^{-8}$, $\beta_{\rm H_{2}O}=1.43$, $\alpha_{\rm H_{2}}=2.572\times 10^{-6}$, and $\beta_{\rm H_{2}}=1.0$ \citep{Lebrun2013, Lichtenberg2021}.
		
\begin{figure}
\gridline{\fig{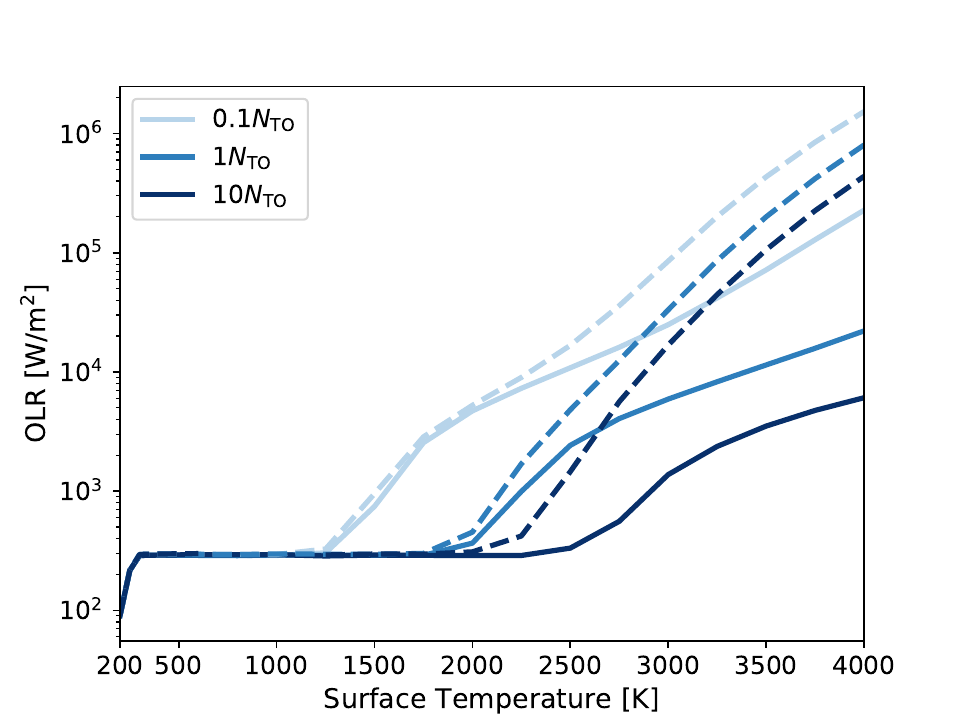}{0.45\textwidth}{(a)}
		\fig{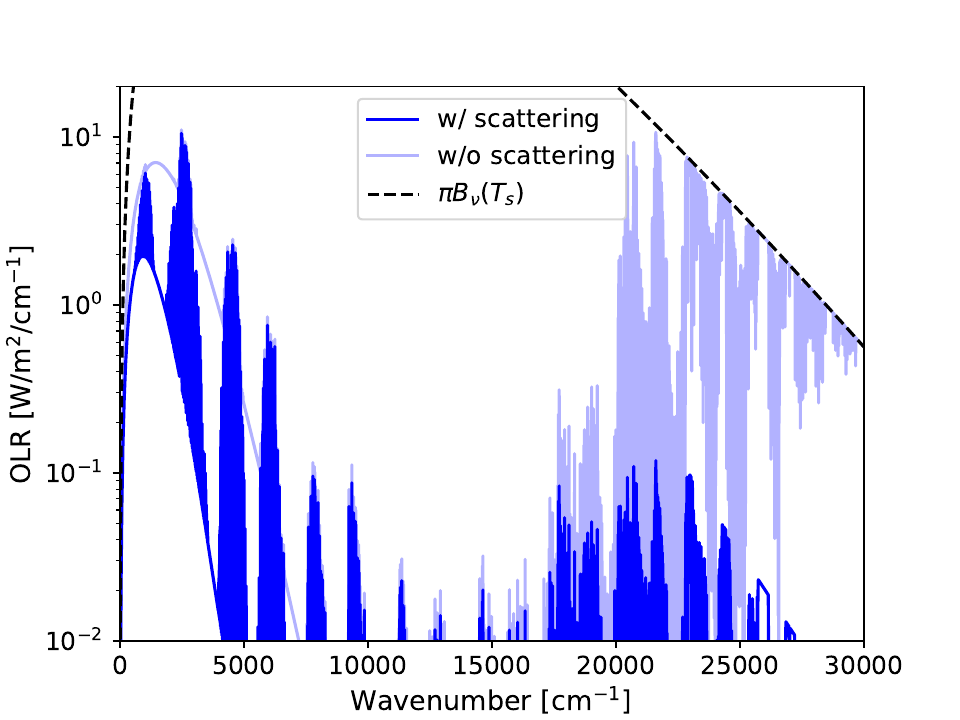}{0.45\textwidth}{(b)}
          	}
\gridline{\fig{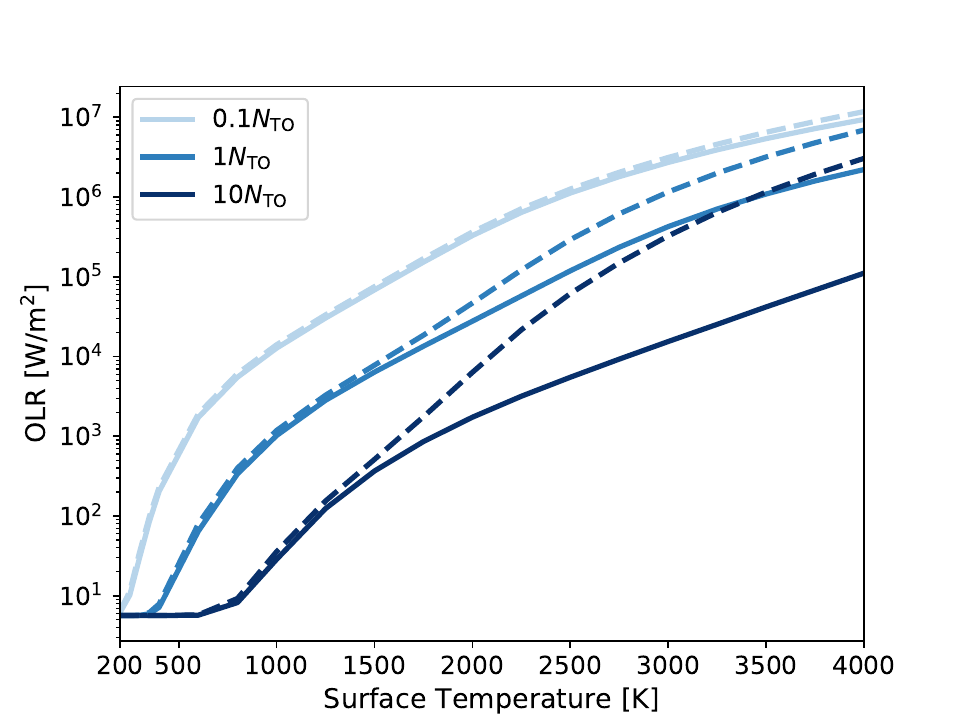}{0.45\textwidth}{(c)}
		\fig{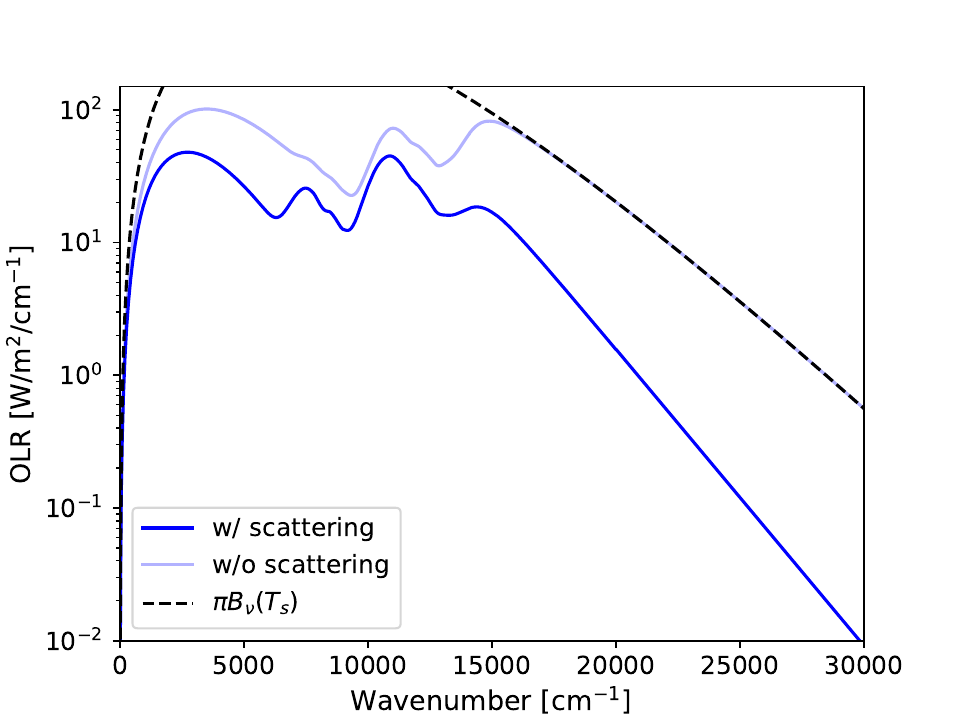}{0.45\textwidth}{(d)}
		}		
\gridline{\fig{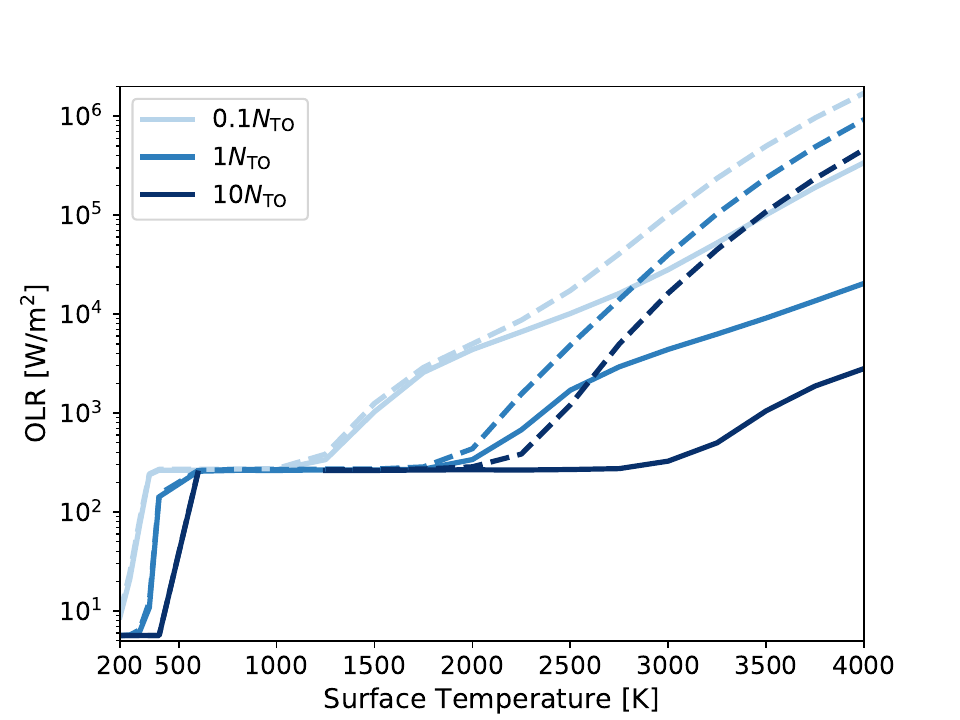}{0.45\textwidth}{(e)}
		\fig{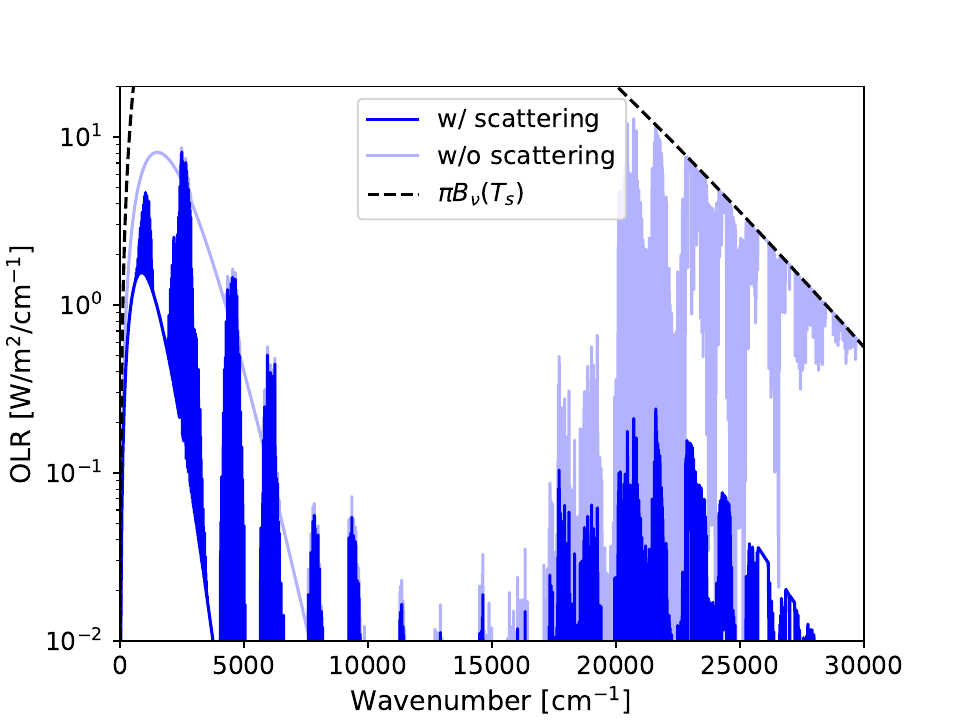}{0.45\textwidth}{(f)}
          	}
\caption{(a) Outgoing longwave radiation (OLR) depending on the surface temperature on pure H$_{2}$O atmospheres. The solid and dashed lines represent the OLR with and without the Rayleigh scattering blanketing effects, respectively. Line colors represent variations in the total hydrogen molecular number in the atmosphere with seawater. Here $N_{\rm TO}$ is the hydrogen molecular number in the present-day terrestrial seawater. (b) OLR per wavenumber when the surface temperature is 3000 K and the total hydrogen amount is $N_{\rm TO}$ on a pure H$_{2}$O atmosphere with seawater. The dark and light blue lines represent the OLR with and without the Rayleigh scattering blanketing effects, respectively. The dashed black line represents the blackbody spectrum with the surface temperature. (c) and (d) are the same as (a) and (b), respectively, but for the pure H$_{2}$ atmospheres. (e) and (f) are the same as (a) and (b), respectively, but for the H$_{2}$O-H$_{2}$ atmospheres with a molar H$_{2}$/H$_{2}$O ratio of unity in the atmosphere with seawater.}
\label{fig:1}
\end{figure}

\begin{figure}
\gridline{\fig{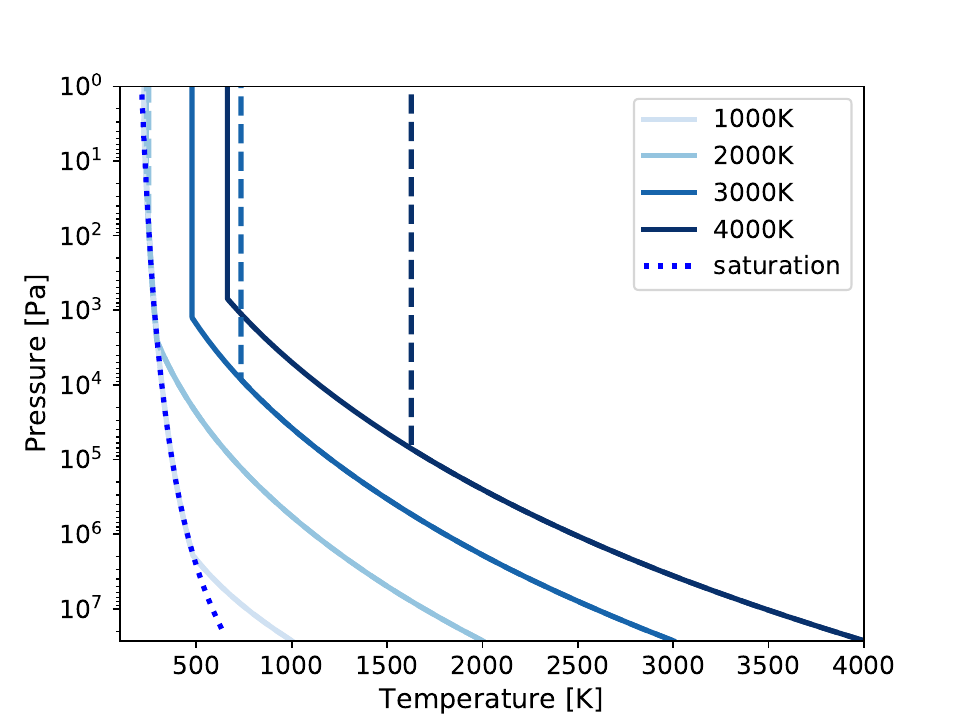}{0.35\textwidth}{(a)}
		\fig{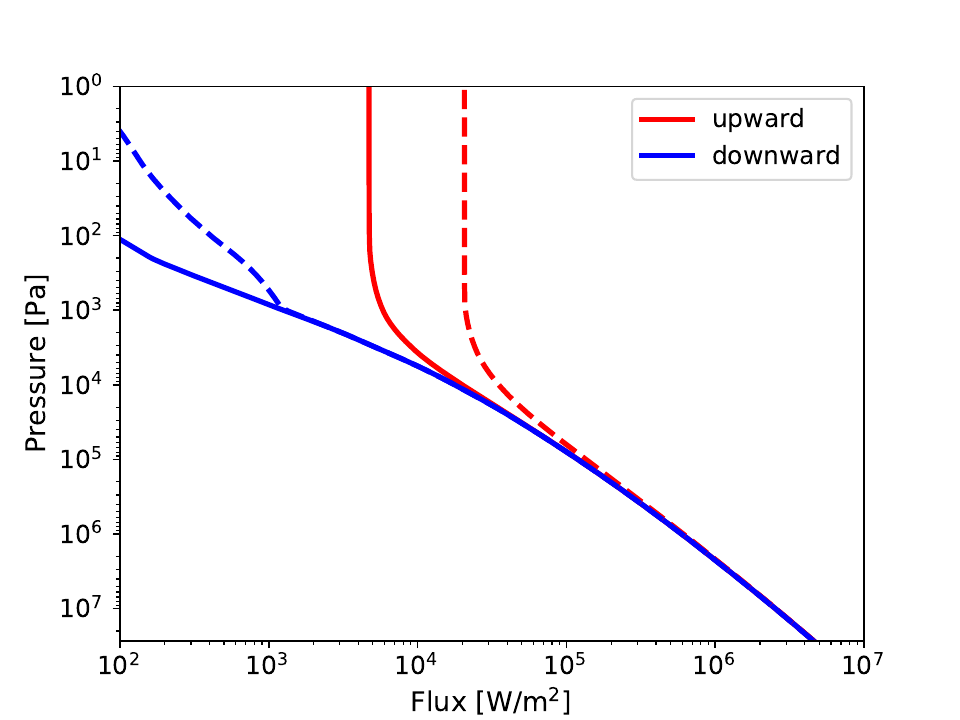}{0.35\textwidth}{(b)}
		\fig{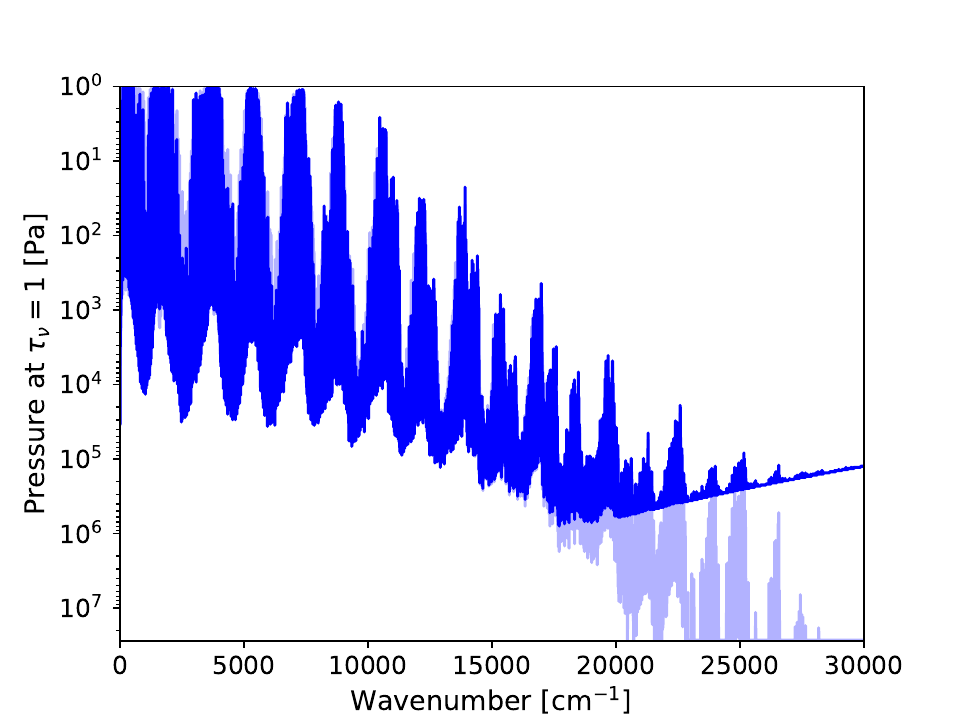}{0.35\textwidth}{(c)}
 	       	}
\gridline{\fig{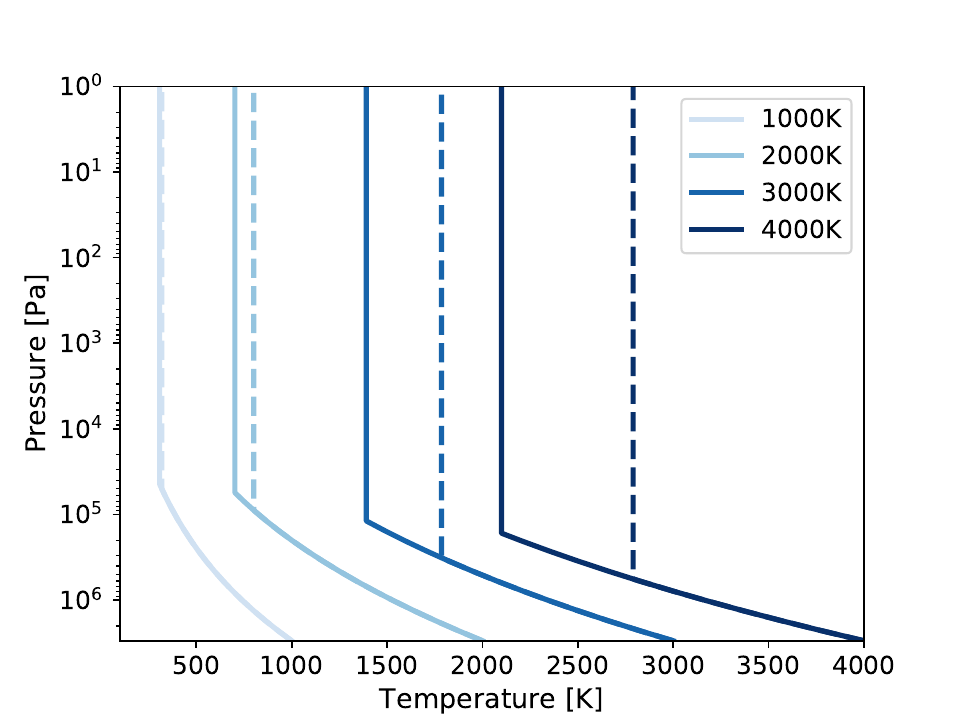}{0.35\textwidth}{(d)}
		\fig{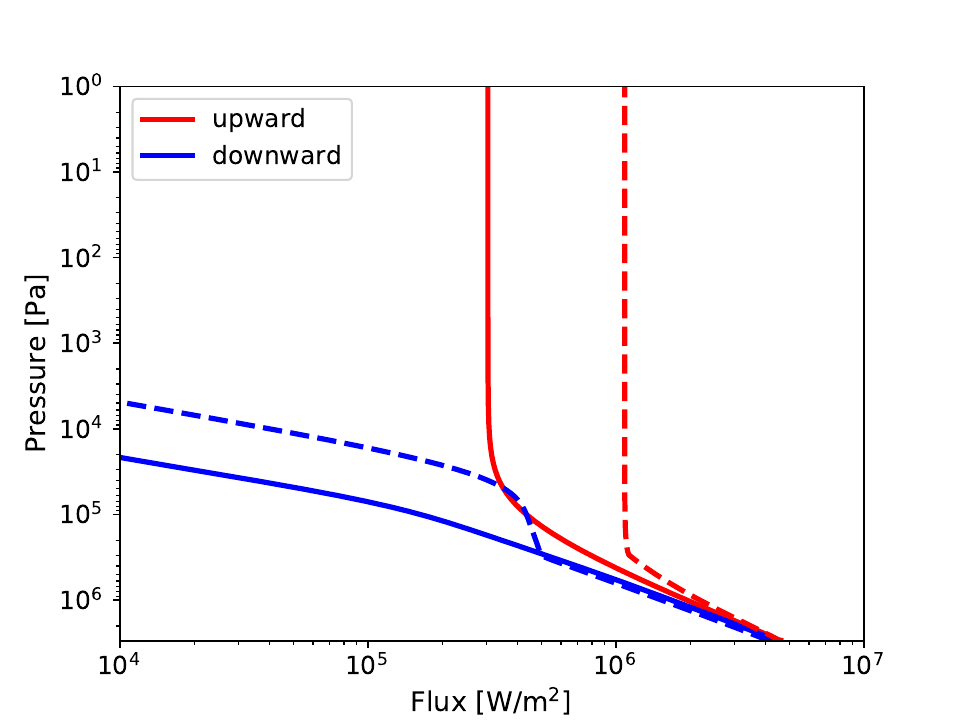}{0.35\textwidth}{(e)}
		\fig{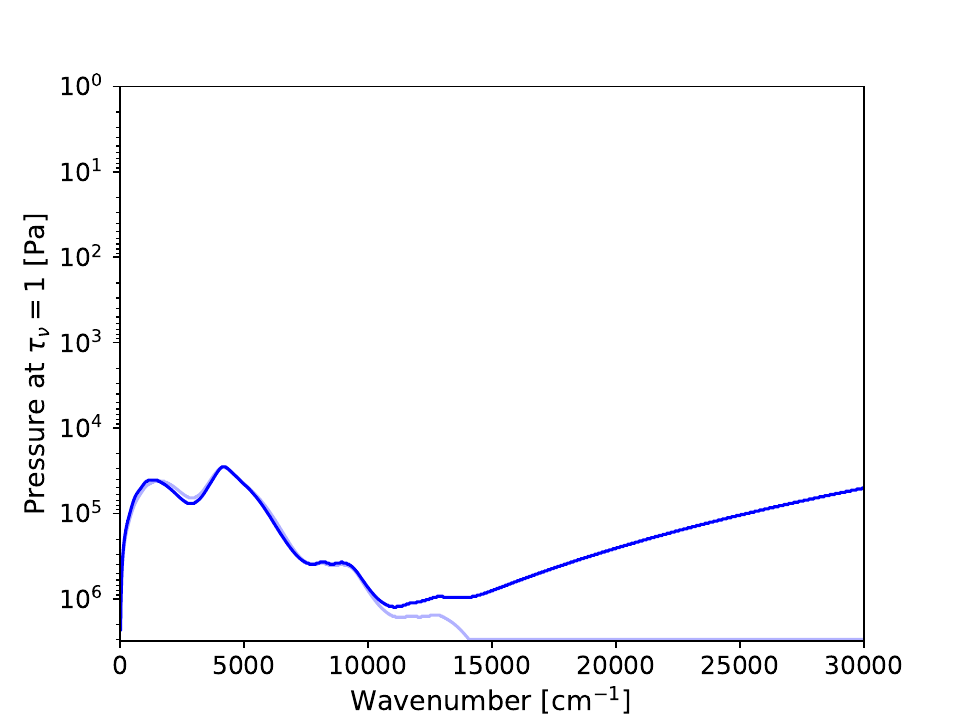}{0.35\textwidth}{(f)}
		}
\gridline{\fig{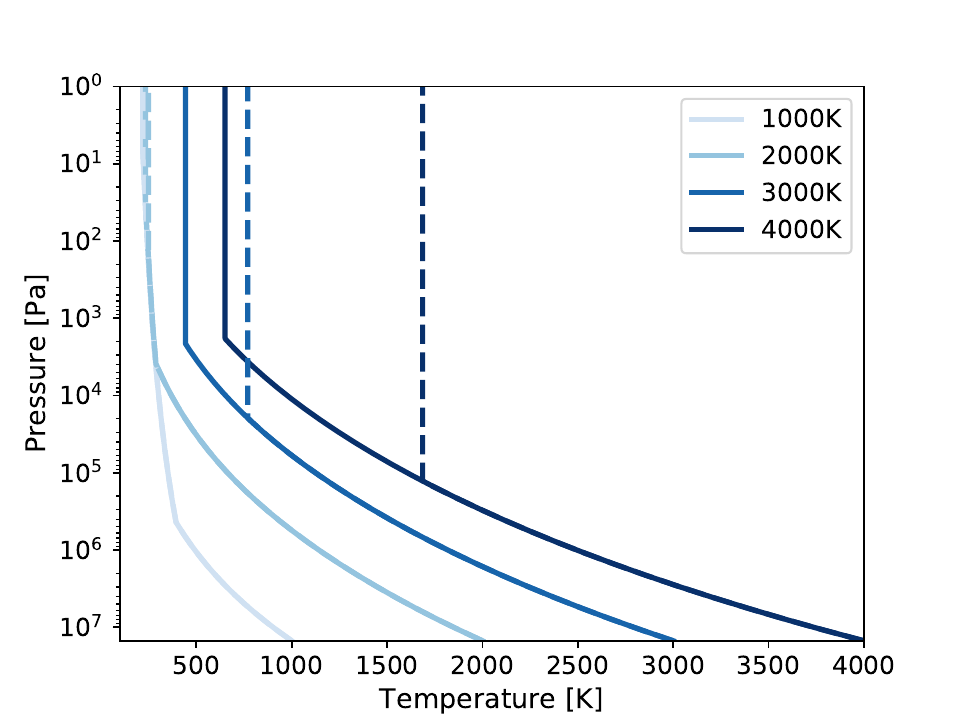}{0.35\textwidth}{(g)}
		\fig{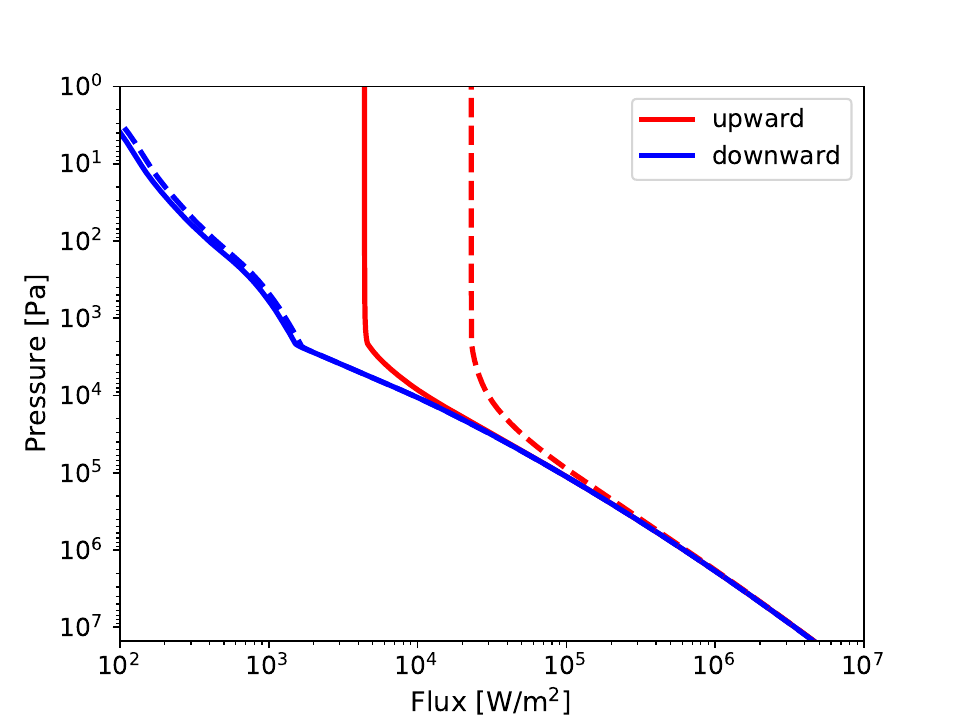}{0.35\textwidth}{(h)}
		\fig{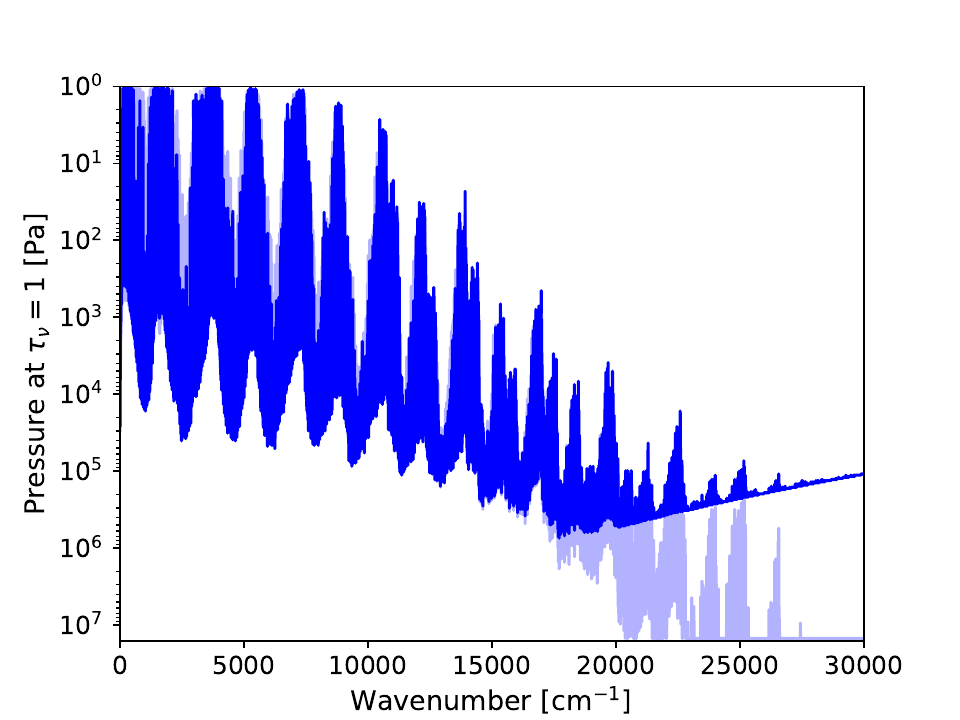}{0.35\textwidth}{(i)}
          	}
\caption{(a) Temperature profile in the pure H$_{2}$O atmosphere with the total hydrogen amount of $N_{\rm TO}$. Here $N_{\rm TO}$ is the molecular number of hydrogen in the present-day terrestrial seawater. The solid and dashed lines represent the profiles with and without the Rayleigh scattering blanketing effects, respectively. Line colors represent variations in the surface temperature. The blue dotted line is the H$_2$O saturation vapor pressure. (b) Profile of the upward and downward planetary radiation flux in the pure H$_{2}$O atmosphere with the total hydrogen amount of $N_{\rm TO}$ and the surface temperature of 3000 K. The red and blue lines represent the upward and downward fluxes, respectively. The solid and dashed lines represent the profiles with and without the Rayleigh scattering blanketing effects, respectively. (c) Pressure where $\tau_{\nu}=1$ with wavenumber in the pure H$_{2}$O atmosphere with the total hydrogen amount of $N_{\rm TO}$ and the surface temperature of 3000 K. The dark and light blue lines represent the values with and without the Rayleigh scattering blanketing effects, respectively. (d), (e), and (f) are the same as (a), (b), and (c), respectively, but for the pure H$_{2}$ atmosphere with the total hydrogen amount of $N_{\rm TO}$. (g), (h),  and (i) are the same as (a), (b), and (c), respectively, but for the H$_{2}$O-H$_{2}$ atmosphere with the molar H$_{2}$/H$_{2}$O ratio of unity and the total hydrogen amount of $N_{\rm TO}$.}
\label{fig:2}
\end{figure}

\begin{figure}
\gridline{\fig{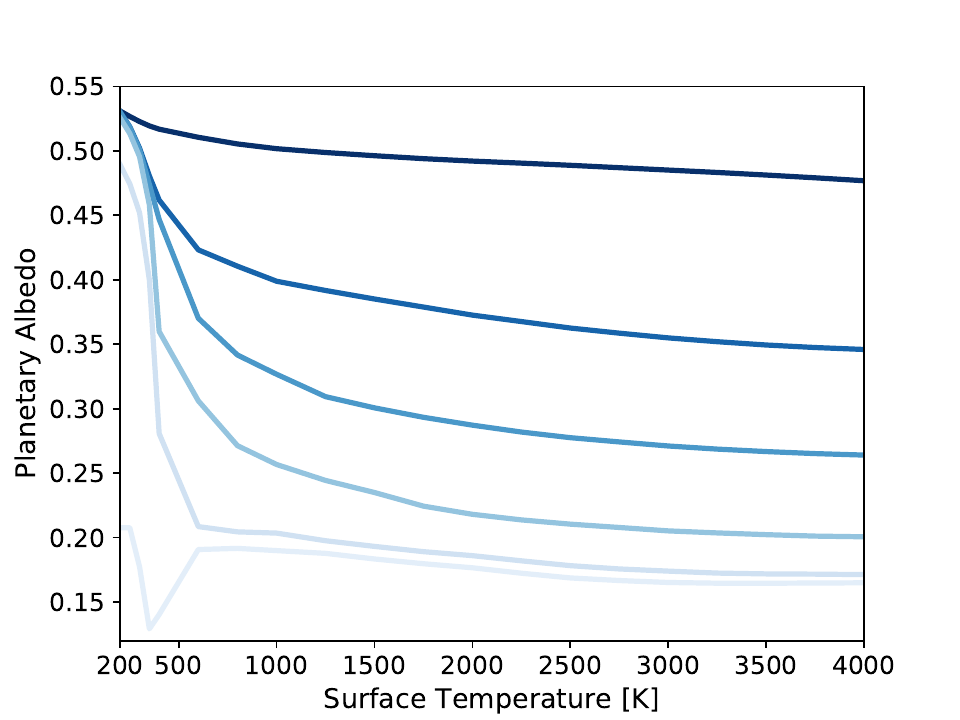}{0.45\textwidth}{(a)}
		\fig{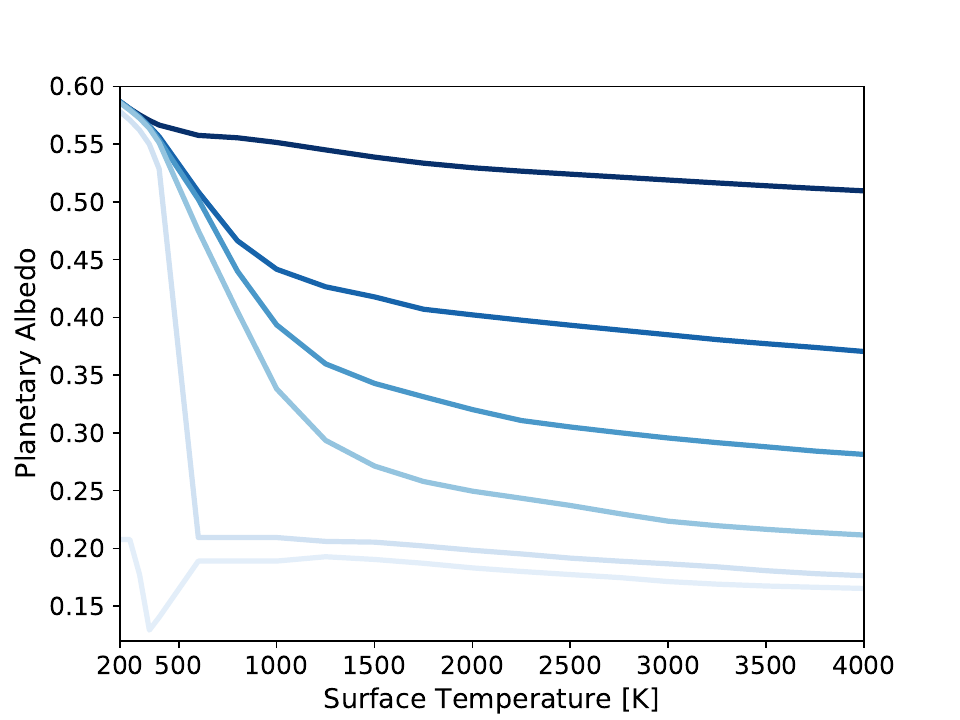}{0.45\textwidth}{(b)}
		}		
\gridline{\fig{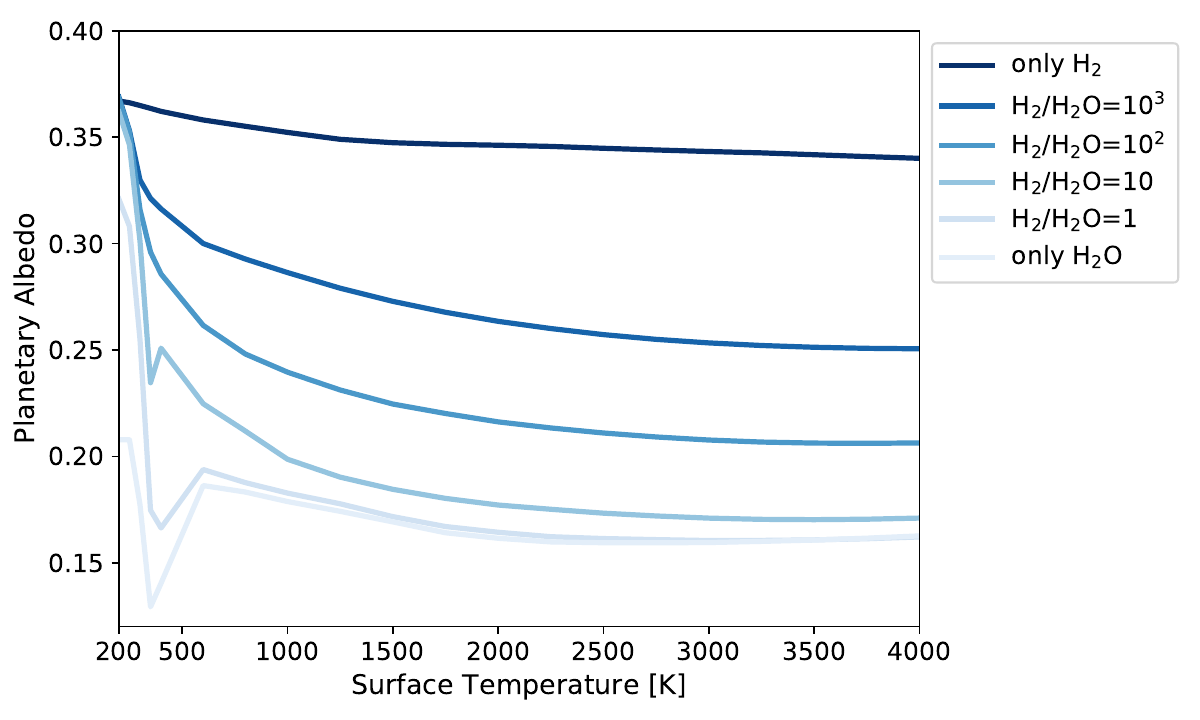}{0.50\textwidth}{(c)}
          	}
\caption{Planetary albedo with surface temperature when the total hydrogen molecular numbers in the atmosphere with seawater are $1N_{\rm TO}$ (a), $10N_{\rm TO}$ (b), and $0.1N_{\rm TO}$ (c), respectively. Here $N_{\rm TO}$ is the hydrogen molecular number in the present-day terrestrial seawater. Line colors show variations in the molar H$_{2}$/H$_{2}$O ratio in the atmosphere with seawater.}
\label{fig:3}
\end{figure}

\begin{figure}
\gridline{\fig{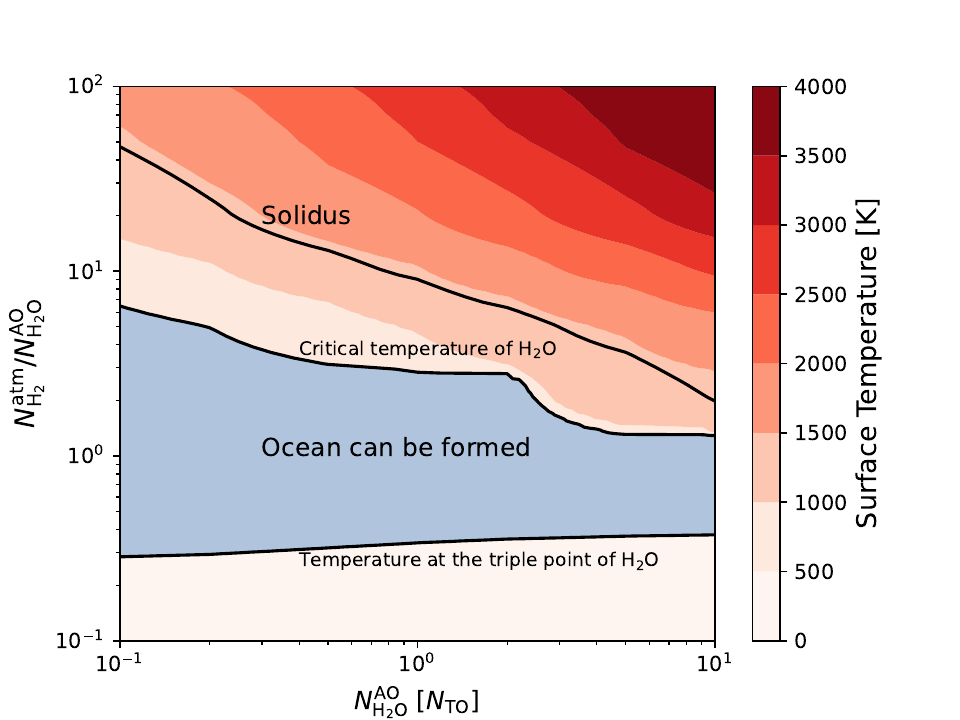}{0.45\textwidth}{(a)}
		\fig{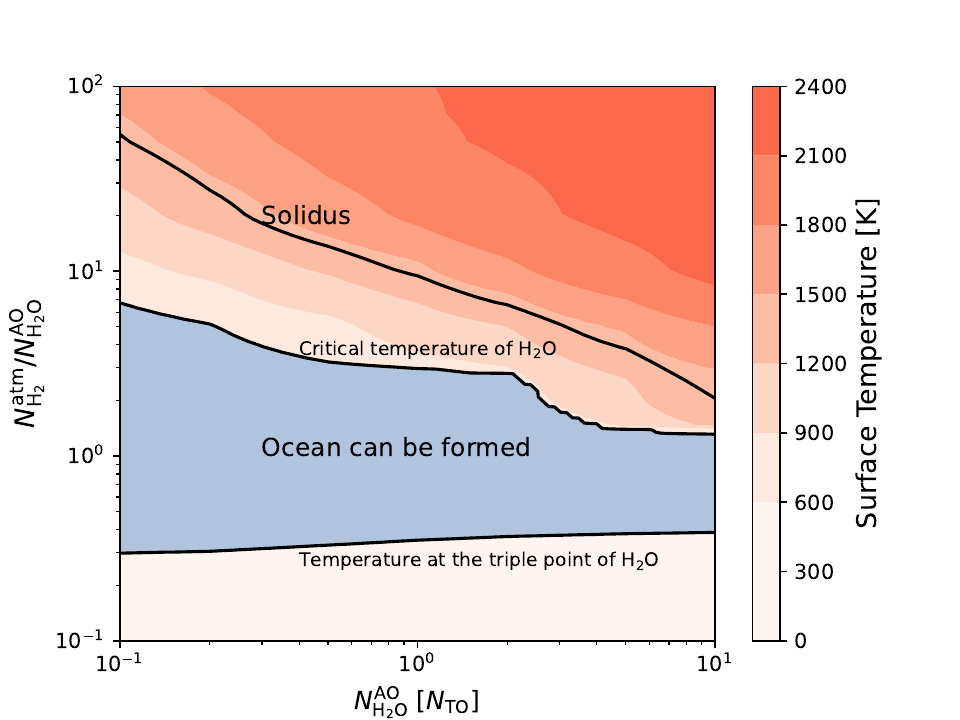}{0.45\textwidth}{(b)}
		}		
\caption{Equilibrium surface temperature where OLR equals net absorbed solar flux as a function of amounts of H$_2$O and H$_2$ on the planetary surface. (a) and (b) are the results with and without the scattering blanketing effects, respectively. The horizontal axis $N_{\rm H_{2}O}^{\rm AO}$ is the hydrogen molecular number of H$_2$O in the atmosphere with seawater normalized by the hydrogen molecular number in the present-day terrestrial seawater ($N_{\rm TO}$). The vertical axis $N_{\rm H_{2}}^{\rm atm}/N_{\rm H_{2}O}^{\rm AO}$is the ratio of the atmospheric hydrogen molecular number of H$_2$ to the hydrogen molecular number of H$_2$O in the atmosphere with seawater. The black lines represent the solidus temperature, the critical temperature of H$_{2}$O, and the temperature at the triple point of H$_{2}$O from the top. In the blue regions, surface temperatures are between the temperature at the triple point and the critical temperature, where oceans can be formed.}
\label{fig:4}
\end{figure}

\begin{figure}[htbp]
	\centering
	\includegraphics[width=0.45\columnwidth]{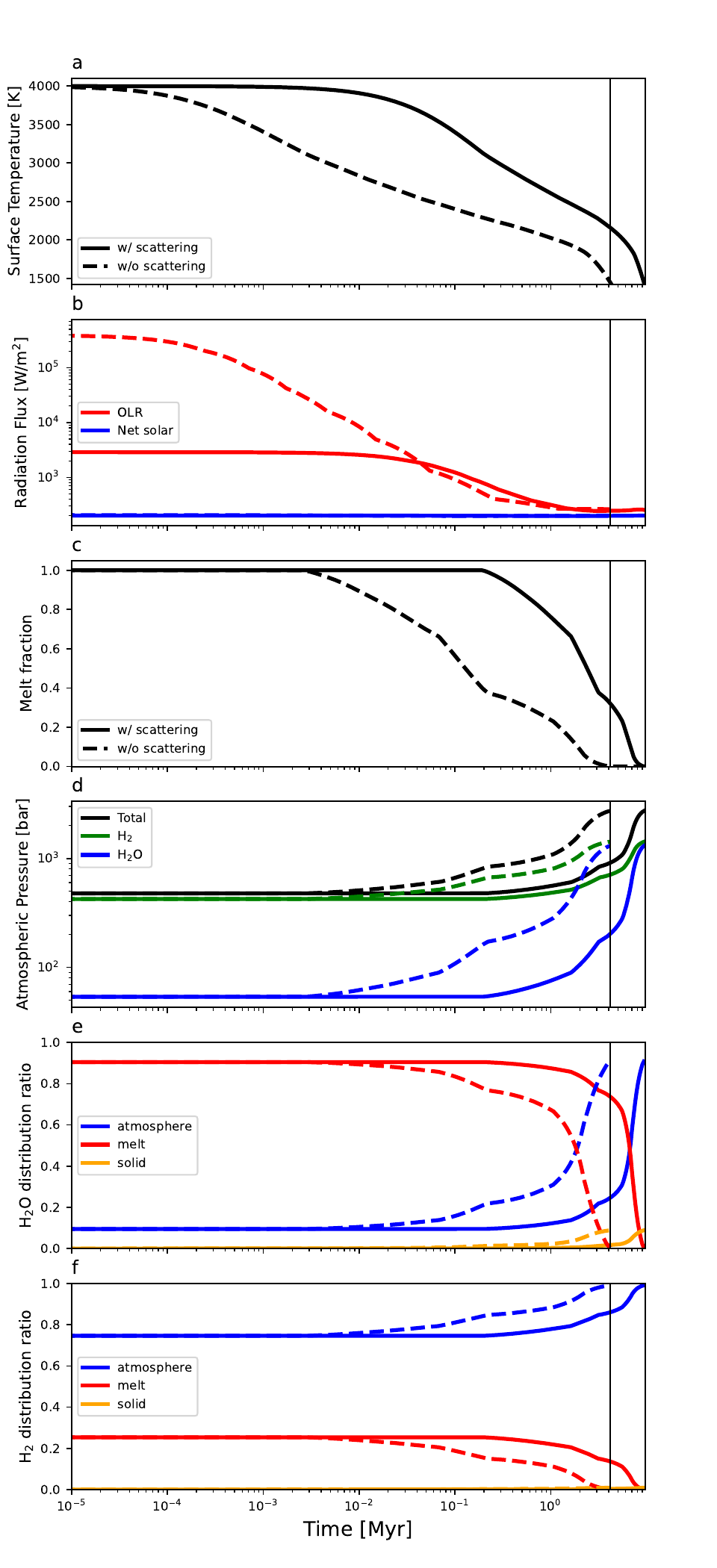}
	\caption{An example of the thermal evolution of magma ocean and atmosphere when the total amounts of both H$_2$O and H$_2$ are $10N_{\rm TO}$. Here $N_{\rm TO}$ is the hydrogen molecular number in the present-day terrestrial seawater. The solid and dashed lines represent the results with and without the scattering blanketing effect, respectively. The vertical black lines represent the time when the magma ocean solidification finished in the case without scattering. (a) Surface temperature with time. (b) Radiation flux with time. The red lines represent OLR and the blue lines represent the net absorbed solar flux. (c) Melt fraction in the whole mantle. (d) Atmospheric pressure with time. The black line, green line, and blue line represent the total surface pressure, partial surface pressure of H$_2$, and partial surface pressure of H$_2$O, respectively. (e) The distribution ratio of H$_2$O among the atmosphere, melt, and solid. The blue, red, and orange lines represent the ratios of the atmosphere, melt, and solid, respectively. (f) The distribution ratio of H$_2$ among the atmosphere, melt, and solid. The blue, red, and orange lines represent the ratios of the atmosphere, melt, and solid, respectively.}
	\label{fig:5}
\end{figure}

\begin{figure}
\gridline{\fig{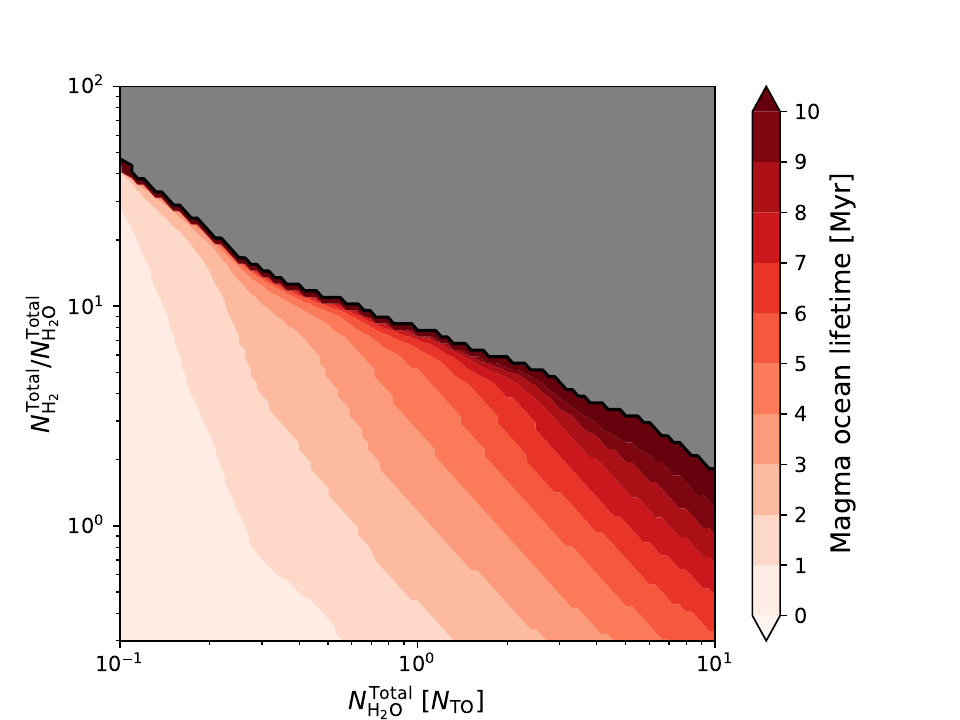}{0.45\textwidth}{(a)}
		\fig{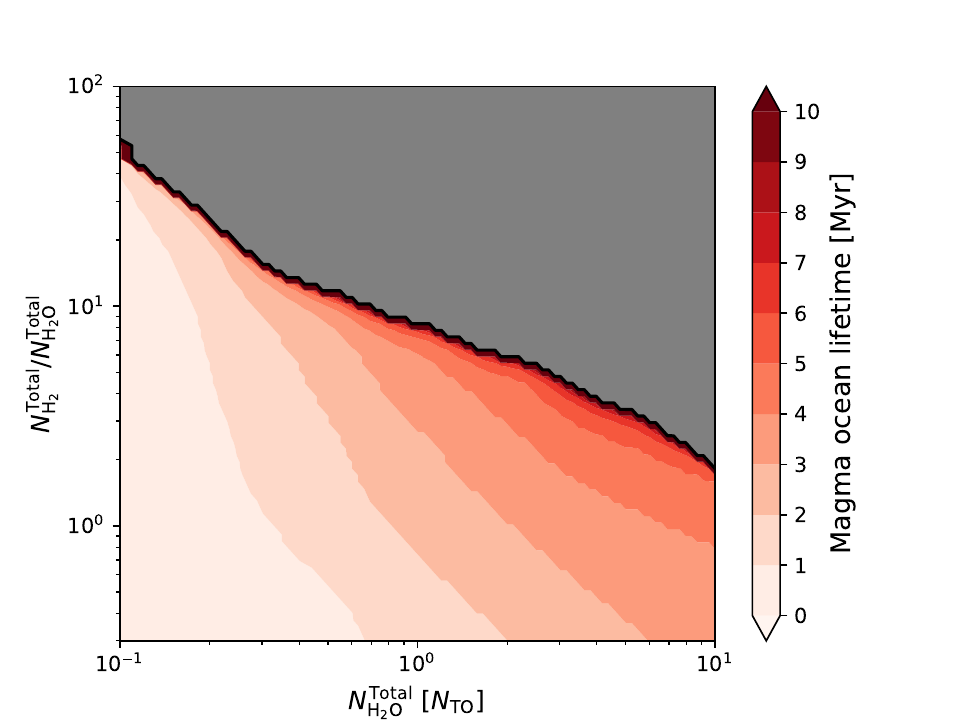}{0.45\textwidth}{(b)}
		}		
\caption{Lifetime of the magma ocean, defined as the duration over which the surface temperature remains above the solidus temperature, as a function of the total amounts of H$_{2}$O and H$_{2}$. (a) and (b) are the results with and without the scattering blanketing effects, respectively. The horizontal axis is the total hydrogen molecular number of H$_2$O normalized by the molecular number of hydrogen in the present-day terrestrial seawater ($N_{\rm TO}$). The vertical axis is the ratio of the total amount of H$_2$ to that of H$_2$O. In the grey region, the equilibrium surface temperature exceeds the solidus temperature.}
\label{fig:6}
\end{figure}

\section{Results} \label{sec:result}
\subsection{Radiative properties and atmospheric structure}
	In this section, we illustrate the radiation properties, with a particular focus on the scattering blanketing effect. This section only considers the atmosphere with condensed seawater on the planetary surface, excluding interactions with the planetary interior described in Section 2.3 and 2.4. Outgoing longwave radiation (OLR) is presented in Figure~\ref{fig:1}, and vertical profiles of temperature, radiation flux, and optical depth are shown in Figure~\ref{fig:2}. For pure H$_2$O atmospheres, the scattering blanketing effect becomes more pronounced as surface temperature and pressure increase, as demonstrated by the differences in OLR between cases with and without scattering (Figure~\ref{fig:1}(a)). Rayleigh scattering suppresses OLR at high wavenumbers ($\gtrsim 20000\,\mathrm{cm^{-1}}$), while H$_2$O absorption dominates at lower wavenumbers (Figure~\ref{fig:1}(b)). At high temperatures, OLR increases at higher wavenumbers where Rayleigh scattering is significant (Equation (16)), enhancing the scattering blanketing effect. Additionally, high pressure is required for the scattering blanketing effect to become significant, as the optical depth at wavenumbers dominated by Rayleigh scattering reaches unity when atmospheric pressure exceeds $\sim 10^{5}\,\mathrm{Pa}$, as shown in the pressure at the optical depth of unity (Figure~\ref{fig:2}(c)). The effect of Rayleigh scattering is also displayed in the vertical profile of the planetary radiation flux (Figure~\ref{fig:2}(b)), where the flux is significantly suppressed in the scattering case compared to the no-scattering case near and above the pressure level where Rayleigh scattering optical depth becomes unity. This suppression of OLR by the scattering results in a temperature drop in the upper atmosphere (Figure~\ref{fig:2}(a)), leading to a reduction in OLR at low wavenumbers below $\sim 20000~\mathrm{cm^{-1}}$ (Figure~\ref{fig:1}(b)). In certain surface temperature ranges, OLR remains nearly constant regardless of surface temperature (Figure 1(a)). This is known as the radiation limit, which occurs when the atmospheric profile and radiative flux become fixed due to H$_{2}$O saturation around the levels where the optical depth is about unity \citep{Nakajima1992}. The H$_2$O saturation is shown in the coincidence of H$_2$O pressure and its saturation vapor pressure under radiation-limit conditions in Figure~\ref{fig:2}(a).
	
	Rayleigh scattering is also effective in atmospheres including H$_2$ (Figures~\ref{fig:1}(c)--(f); Figures~\ref{fig:2}(d)--(i)). In pure H$_2$ atmospheres, OLR is suppressed through the H$_{2}$-H$_{2}$ collision-induced absorption at wavenumbers below $\sim15000\,\mathrm{cm^{-1}}$, as well as through Rayleigh scattering (Figures~\ref{fig:1}(c) and (d); Figures~\ref{fig:2}(e) and (f)). Compared to pure H$_2$O atmosphere cases, the pressure level for the optical depth to reach unity tends to be high (Figures~\ref{fig:2}(e) and (f)). In the case when the H$_2$/H$_2$O ratio is unity, both the H$_2$O absorption and H$_2$-H$_2$ collision-induced absorption work effectively along with Rayleigh scattering (Figure~\ref{fig:1}(f); Figures~\ref{fig:2}(h) and (i)), leading to the lower OLR than that of the pure H$_2$ cases (Figure~\ref{fig:1}(e)).

	The planetary albedo depending on the surface temperature, atmospheric mass, and composition is shown in Figure~\ref{fig:3}. The planetary albedo is strongly influenced by the atmospheric H$_2$/H$_2$O ratio, increasing with higher ratios due to a decrease in stellar absorption by H$_2$O under H$_2$-rich conditions. Higher total atmospheric mass also enhances the albedo through more effective scattering at high pressures, as shown in the comparisons among Figure~\ref{fig:3}(a), (b), and (c). The gradual decreases in planetary albedo with increasing surface temperature are driven by an increased atmospheric H$_2$O abundance, broadening of H$_2$O absorption lines \citep[e.g.,][]{Rothman2010}, and the enhanced absorption coefficient of the H$_2$-H$_2$ collision-induced absorption at higher temperatures \citep[e.g.,][]{Borysow2002}. Although this study assumes solar-type stellar flux, planetary albedo also depends on the stellar type. In particular, albedo is expected to significantly decrease for cooler stars, as stellar emission shifts to longer wavelengths, leading to stronger atmospheric absorption and weaker Rayleigh scattering of stellar irradiation \citep{Kopparapu2013, Pluriel2019}.
	
	To clarify the fundamental relationship between planetary and solar radiation, equilibrium surface temperature, where OLR as shown in Figure~\ref{fig:1} equals net absorbed solar flux derived from planetary albedo (Figure~\ref{fig:3}) and Equation (19) under fixed atmospheric mass and composition, is shown in Figures~\ref{fig:4}. The equilibrium surface temperature increases as the amounts of H$_{2}$O and H$_{2}$ increase due mainly to the suppression of OLR. The scattering blanketing effect further enhances this OLR suppression, leading to higher surface temperatures, particularly when H$_{2}$O and H$_{2}$ are abundant, as shown in the comparison between Figures~\ref{fig:4}(a) and \ref{fig:4}(b). Consequently, the equilibrium surface temperature exceeds the solidus temperature at high atmospheric masses and H$_2$/H$_2$O ratios. For $\mathrm{H_2}/\mathrm{H_{2}O}\sim 1$, the equilibrium surface temperature lies between the triple point and the critical temperature of H$_{2}$O, suggesting that oceans can be formed under such conditions. Although a parameter space where oceans can exist appears under the solar flux at Earth's orbit, this would vanish under much higher stellar fluxes exceeding the radiation limit, as in the case of Venus's orbit. As the H$_2$/H$_2$O ratio decreases, the equilibrium surface temperature becomes less sensitive to total atmospheric mass and more influenced by the H$_2$/H$_2$O ratio. This occurs partly because the increase in planetary albedo associated with the increase in atmospheric amount, which leads to a decrease in surface temperature, offsets the decrease in OLR that contributes to temperature increase. In H$_2$-depleted conditions, the equilibrium surface temperature little depends on the total H$_2$O amount because the atmospheric profile is determined by the saturation vapor pressure.
	
\subsection{Thermal evolution of magma ocean and atmosphere}
	In this section, we estimate the thermal co-evolution of the magma ocean and atmosphere by applying the calculated planetary and solar radiation fluxes as functions of surface temperature, atmospheric amount, and composition. The total amounts of H$_{2}$O and H$_{2}$ are treated as free parameters. As described in Section 2.3 in detail, the cooling and solidification of the magma ocean is supposed to be along with the net energy balance of the incoming and outgoing radiation at the top of the atmosphere. We also consider the partitioning of volatiles between the atmosphere and interior as described in Section 2.4. For simplicity, we neglect atmospheric escape to space, chemical interaction between the magma ocean and atmosphere, and the accretion of materials that could supply volatiles or induce chemical reactions. We stop the evolution calculations either when the surface temperature drops below the solidus temperature or when the integration time exceeds 50 Myr. We define the magma ocean lifetime as the duration over which the surface temperature remains above the solidus temperature.
	
	Figure~\ref{fig:5} presents an example of the thermal evolution of a magma ocean and atmosphere when the total amounts of both H$_2$O and H$_2$ are $10N_{\rm TO}$. In the early stages, the surface temperature is high and the atmosphere is H$_2$-dominated due to more efficient partitioning of H$_2$O into the melt as described in Equation (25) and its Henrian coefficient (Figures~\ref{fig:5}(a), \ref{fig:5}(d)--(f)). During this stage, the scattering blanketing effect significantly suppresses OLR (Figure~\ref{fig:5}(b)). Consequently, the magma ocean maintains relatively high-temperature and high-melt fraction with the scattering blanketing effect (Figures~\ref{fig:5}(a) and \ref{fig:5}(c)). As the magma ocean cools and solidifies via planetary radiation, H$_2$O degassing from the magma ocean progresses (Figures~\ref{fig:5}(d) and \ref{fig:5}(f)), further contributing to OLR suppression (Figure~\ref{fig:5}(b)). The magma ocean lifetime is 9.7 Myr with the scattering blanketing effect, compared to 4.2 Myr without it.
	
	Magma ocean lifetime as a function of total amounts of H$_{2}$O and H$_2$ are shown in Figure~\ref{fig:6}. The lifetime naturally extends as the total amounts of H$_{2}$O and H$_2$ increase, primarily due to the suppression of OLR. As shown in the comparison between Figures~\ref{fig:6}(a) and \ref{fig:6}(b), the scattering blanketing effect on the magma ocean lifetime becomes increasingly pronounced with larger total amounts of H$_2$O and H$_2$, extending the lifetime by up to about three times. In the grey region of Figure~\ref{fig:6}, equilibrium between OLR and absorbed solar flux is achieved at certain stages, sustaining the equilibrium surface temperature above the solidus temperature. This indicates that the magma ocean state can persist as long as the atmospheric mass and composition remain unchanged by processes such as atmospheric escape or chemical interaction with the magma ocean in the ranges of total H$_2$O and H$_2$ amounts.

\section{Discussion} \label{sec:discussion}
\subsection{Effects of uncertainties in assumptions}
\subsubsection{Upper atmospheric temperature}
\begin{figure}
\gridline{\fig{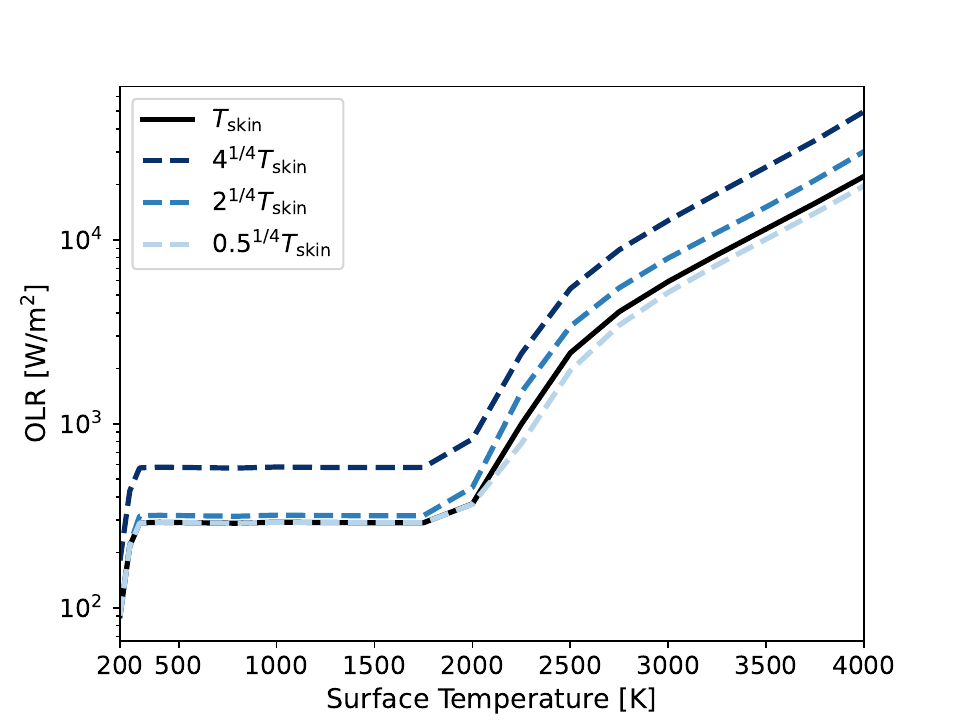}{0.35\textwidth}{(a)}
		\fig{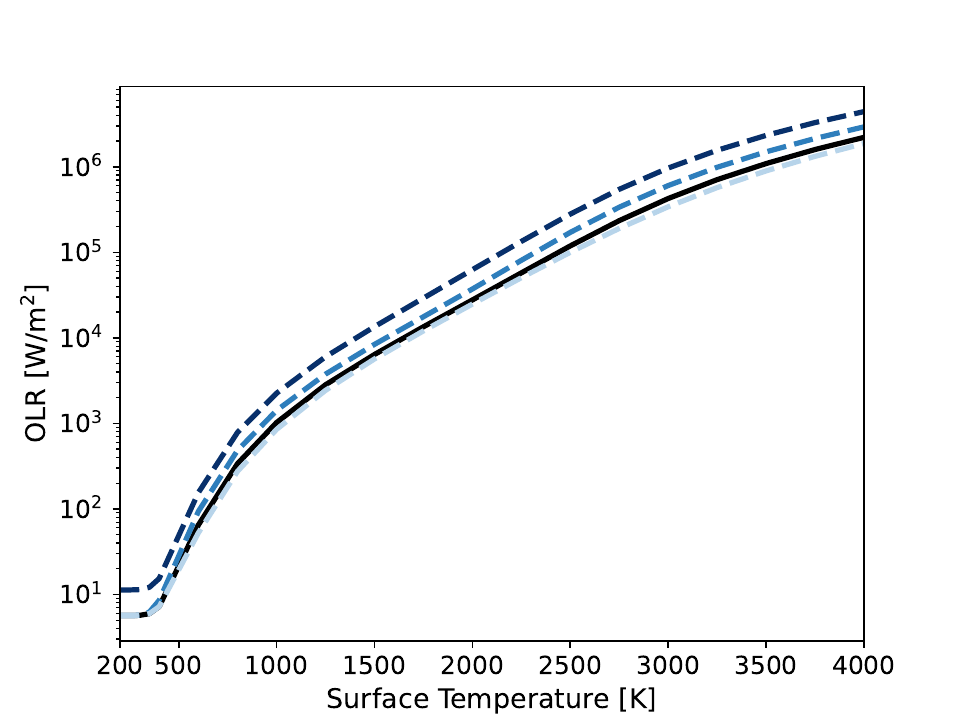}{0.35\textwidth}{(b)}
		\fig{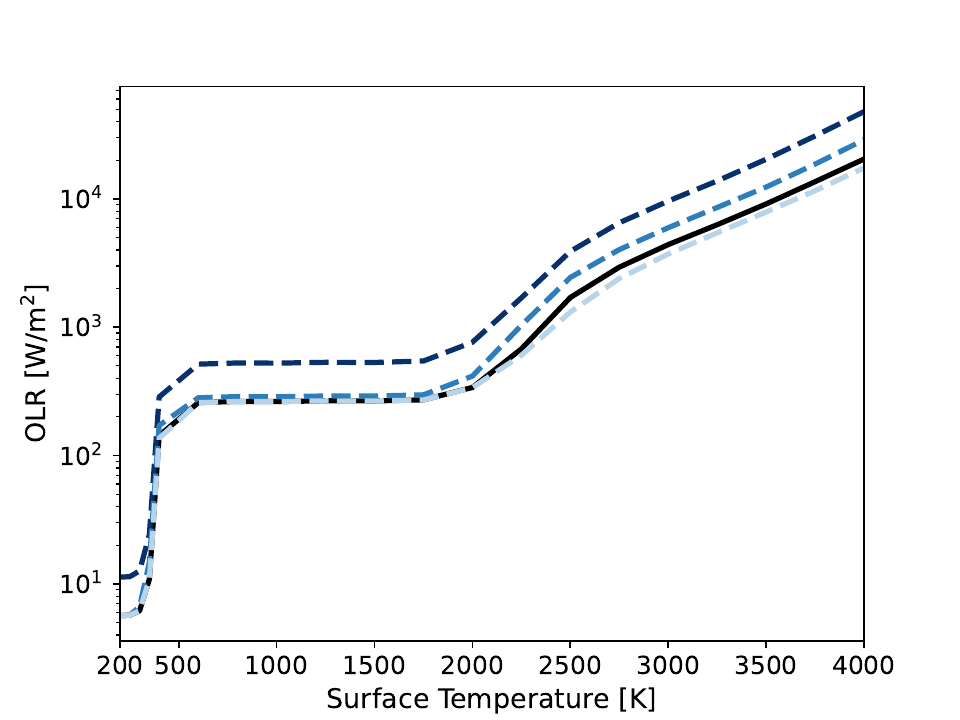}{0.35\textwidth}{(c)}
 	       	}
\caption{Outgoing longwave radiation (OLR) as a function of surface temperature with different upper atmospheric temperatures, for a pure H$_2$O atmosphere (a), a pure H$_2$ atmosphere (b), and an H$_2$O-H$_2$ atmosphere with a molar H$_2$/H$_2$O ratio of unity (c). The black lines represent results with the upper atmospheric temperature set to the skin temperature $T_{\rm skin}$, while the dashed lines represent results for other upper atmospheric temperatures deviating from the skin temperature. In all cases, the atmospheric total hydrogen amount is $N_{\rm TO}$, corresponding to the molecular number of hydrogen in the present-day terrestrial seawater.}
\label{fig:7}
\end{figure}

	Although the upper atmospheric temperature is set equal to the skin temperature in our model, it can deviate from the skin temperature in non-grey atmospheres \citep[e.g.,][]{Wordsworth2013, Leconte2013}. Figure~\ref{fig:7} illustrates the effects of variations in upper atmospheric temperature on OLR. Overall, OLR is largely insensitive to the assumed upper atmospheric temperature because the upper atmosphere is optically thin. However, when the upper atmospheric temperature becomes significantly higher than the skin temperature, OLR increases, as this deviation affects the temperature profile near the level where the optical depth reaches unity. Previous non-grey radiative transfer calculations for H$_2$O-rich atmospheres have shown that the upper atmospheric temperature tends to be lower than the skin temperature because the upper atmosphere can absorb upwelling infrared radiation only in limited spectral regions while remaining an efficient emitter \citep{Wordsworth2013, Leconte2013}. In light of this behavior, variations in the upper atmospheric temperature are expected to have little impact on OLR and the magma ocean lifetime in our calculations.

\subsubsection{Fully convective atmospheric structure}
\begin{figure}
\gridline{\fig{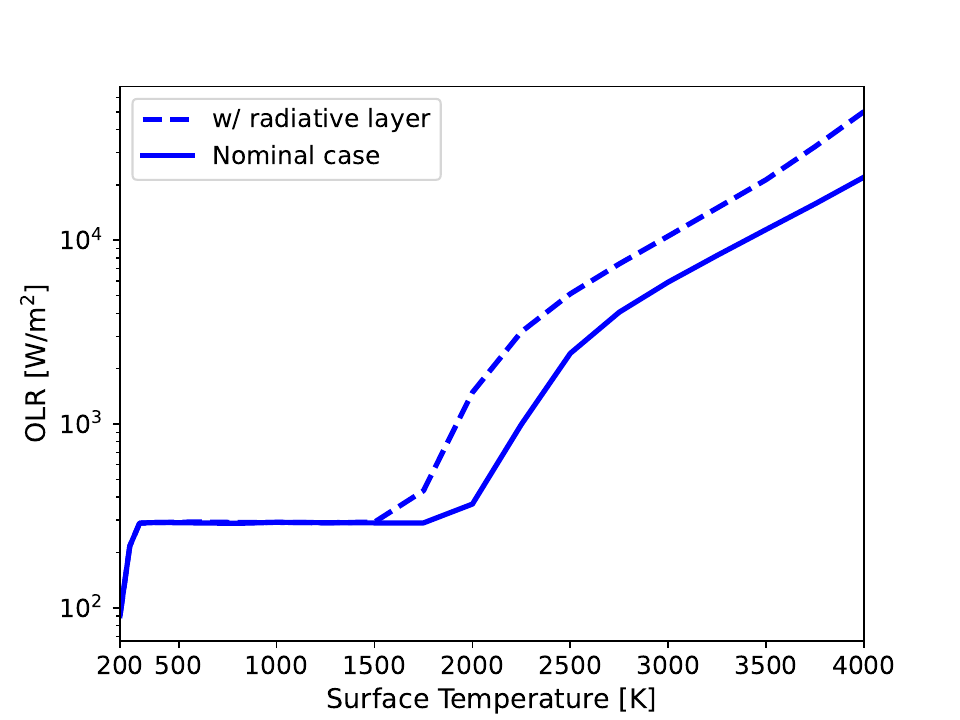}{0.35\textwidth}{(a)}
		\fig{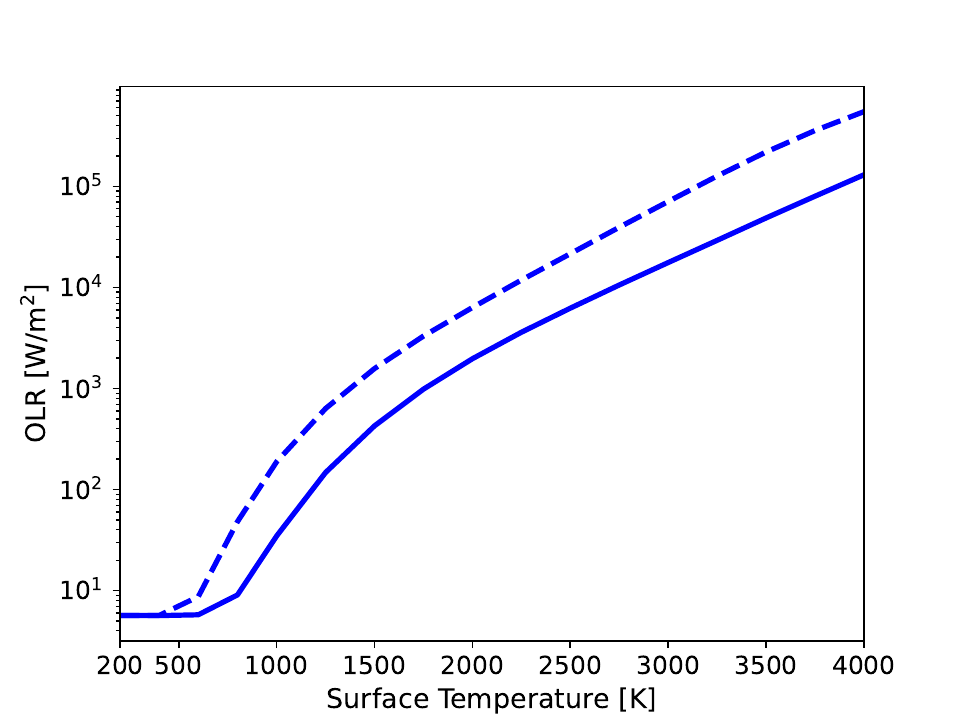}{0.35\textwidth}{(b)}
		\fig{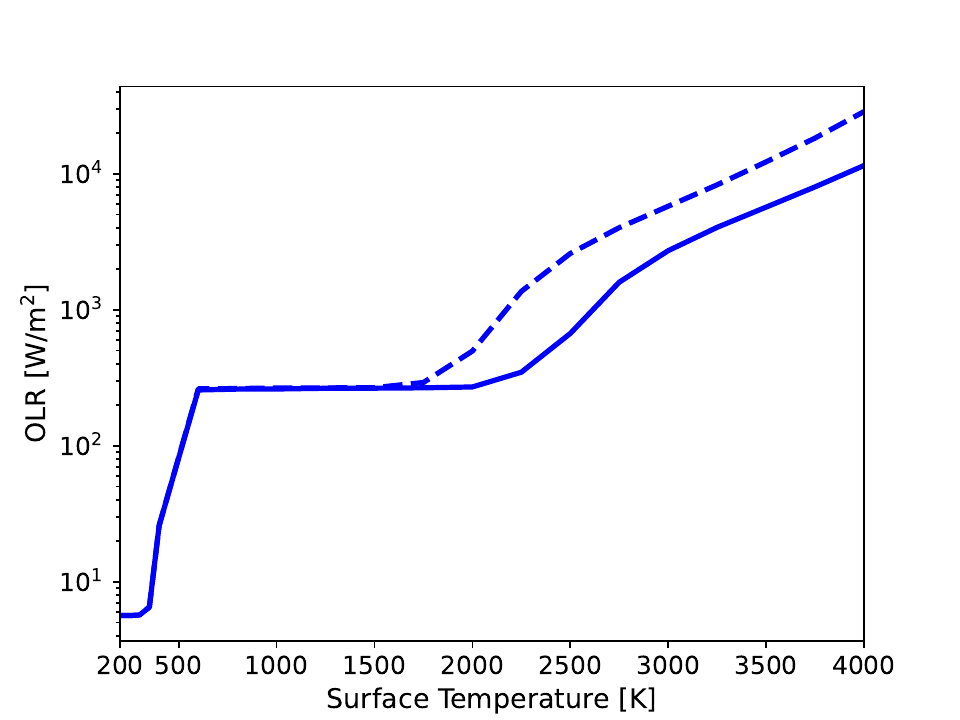}{0.35\textwidth}{(c)}
 	       	}
\caption{Outgonig longwave radiation (OLR) as a function of surface temperature for cases with an isothermal radiative layer at pressures greater than 100 bar (dashed), compared to cases without the radiative layer (solid; standard case), shown for a pure H$_2$O atmosphere (a), a pure H$_2$ atmosphere (b), and an H$_2$O-H$_2$ atmosphere with a molar H$_2$/H$_2$O ratio of unity (c). In all cases, the surface pressure is set to 270 bar, corresponding to the total atmospheric mass equivalent to that of the present-day terrestrial seawater.}
\label{fig:8}
\end{figure}

	Although we assumed that the atmosphere is fully convective below the stratosphere following the majority of previous studies, \citet{Selsis2023} and \citet{Nicholls2025} show that thick proto-atmospheres can become stable against convection and develop a nearly isothermal radiative layer in high-pressure regions due to the absorption of most stellar irradiation in the upper atmosphere, under energy equilibrium condition between planetary and stellar radiation. Figure~\ref{fig:8} presents the effects of such radiative layer formation at pressures greater than 100 bar on OLR. OLR becomes higher by a factor in cases with a deep isothermal layer when the temperature near the level where the optical depth reaches unity increases. Detail investigations of convective stability under extremely high internal heat flux, as considered in this study, are important aspects to be explored in future studies.
	
	Furthermore, H$_2$O condensation in H$_2$-rich atmospheres can also suppress convection by forming a gradient in the atmospheric mean molecular weight, leading to the formation of a super-adiabatic radiative layer in deep regions \citep{Leconte2017, Leconte2024}. This convective inhibition can lower the temperature at and above the super-adiabatic radiative layer for a given surface temperature, thereby reducing OLR and prolonging the magma ocean lifetime. Investigating this condensation-induced convective inhibition is also part of our key future work for understanding the evolution of magma oceans and surface conditions under H$_2$-rich atmospheres.
	
\subsubsection{Cloud-free atmosphere}
\begin{figure}[htbp]
	\centering
	\includegraphics[width=0.45\columnwidth]{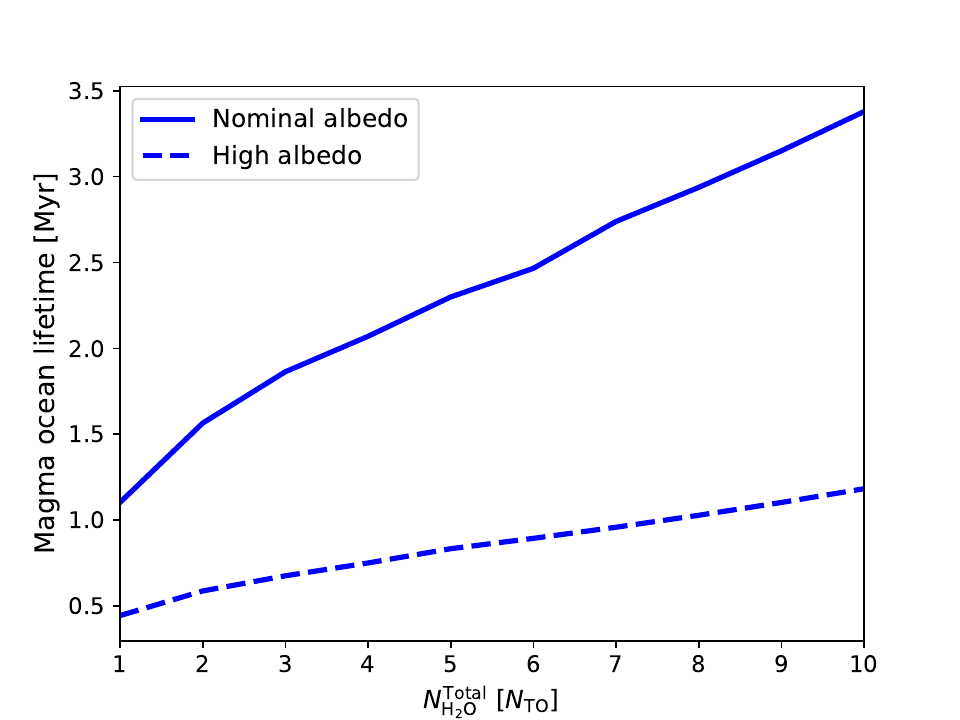}
	\caption{Lifetime of the magma ocean, defined as the duration over which the surface temperature remains above the solidus temperature, under an H$_2$O atmosphere with a nominal planetary albedo of $\sim 0.2$, derived from our model and shown in Figure~\ref{fig:3} (solid line), and a fixed high planetary albedo of 0.9 (dashed line). The horizontal axis is the total hydrogen molecular number of H$_2$O normalized by the molecular number of hydrogen in the present-day terrestrial seawater ($N_{\rm TO}$).}
	\label{fig:9}
\end{figure}

	Although our 1-D model neglects the radiative effects of clouds for simplicity, clouds can significantly influence the atmospheric thermal structure and radiative energy balance \citep{Leconte2013, Charnay2013, Pluriel2019}. In particular, they can substantially enhance the planetary albedo. To assess the potential impact, Figure~\ref{fig:9} compares the magma ocean lifetime under an H$_2$O atmosphere with a fixed high planetary albedo of 0.9, the upper limit estimated by \citet{Pluriel2019}, to that in the nominal case with a planetary albedo of $\sim 0.2$, derived from our model and shown in Figure~\ref{fig:3}. In the high-albedo case, the magma ocean lifetime is reduced by up to about a factor of three due to the less efficient absorption of incoming stellar radiation. The actual lifetime is expected to fall between the values obtained in the nominal and high-albedo cases. A more detailed investigation of cloud effects is therefore crucial for accurately determining the magma ocean lifetime.

\subsection{Chemical evolution of a proto-atmosphere and magma ocean}
	Although this study simplifies the chemical evolution in a proto-atmosphere and magma ocean to focus on the blanketing effect and thermal evolution, it is important to note that the redox states of both the magma ocean and the proto-atmosphere may have changed dramatically over time. During core-mantle differentiation, the magma ocean was likely highly reducing, with a redox state below the iron-w\"{u}stite buffer \citep[e.g.,][]{Frost2008}. Assuming chemical equilibrium with the magma ocean, the proto-atmosphere would have mainly consisted of reduced species such as H$_2$, CH$_4$, and CO \citep[e.g.,][]{Kuramoto1996, Hirschmann2012}. Following the completion of core-mantle differentiation, the redox disproportionation of ferrous iron (Fe$^{2+}$) to ferric iron (Fe$^{3+}$) and metallic iron may have oxidized the magma ocean to the redox level observed in Earth's mantle today, through the sequestration of metallic iron into the core and the homogenization of iron oxides via vigorous convection \citep[e.g.,][]{Armstrong2019, Deng2020, Kuwahara2023, Zhang2024}. Interaction with this oxidized magma ocean would have further oxidized the proto-atmosphere, such as conversion of H$_2$ into H$_2$O.

	Hydrodynamic escape, driven by intense X-ray and extreme ultraviolet (XUV) irradiation from the young Sun, likely also altered the composition of the proto-atmosphere. In H$_2$-rich atmospheres containing carbon-bearing species and H$_2$O, H$_2$ is estimated to escape at rates $\lesssim 1\,\mathrm{bar/Myr}$ while other heavier species largely remain, due to radiative cooling of the outflow by radiatively active species, under early solar XUV conditions \citep{Yoshida2021, Yoshida2022, Yoshida2024b}. Although the rate is estimated to be slow, the preferential loss of H$_2$ can enhance the magma ocean cooling.
	
	Photochemical processes likely also impacted proto-atmospheric composition. Reduced species such as CH$_4$ and CO should be converted into more oxidized species and various organics in the upper atmosphere, where molecular photolysis occurs \citep[e.g.,][]{Zahnle2020, Wogan2023, Yoshida2024a}. However, these photochemical products may have decomposed in the hot, lower atmosphere while the magma ocean remained. The accumulation of photochemical products likely began after the surface temperature dropped enough to allow surface ocean formation. At this stage, CH$_4$ photolysis would primarily produce heavier organics due to their UV self-shielding effects, which suppress the production of oxidant radicals such as OH followed by the H$_2$O photolysis, and the accumulation of organics on the surface, potentially linked to the emergence of living organisms, may have proceeded \citep{Yoshida2024a}.

\subsection{Effects of the prolonged magma ocean lifetime on chemical differentiation}
\begin{figure}
\gridline{\fig{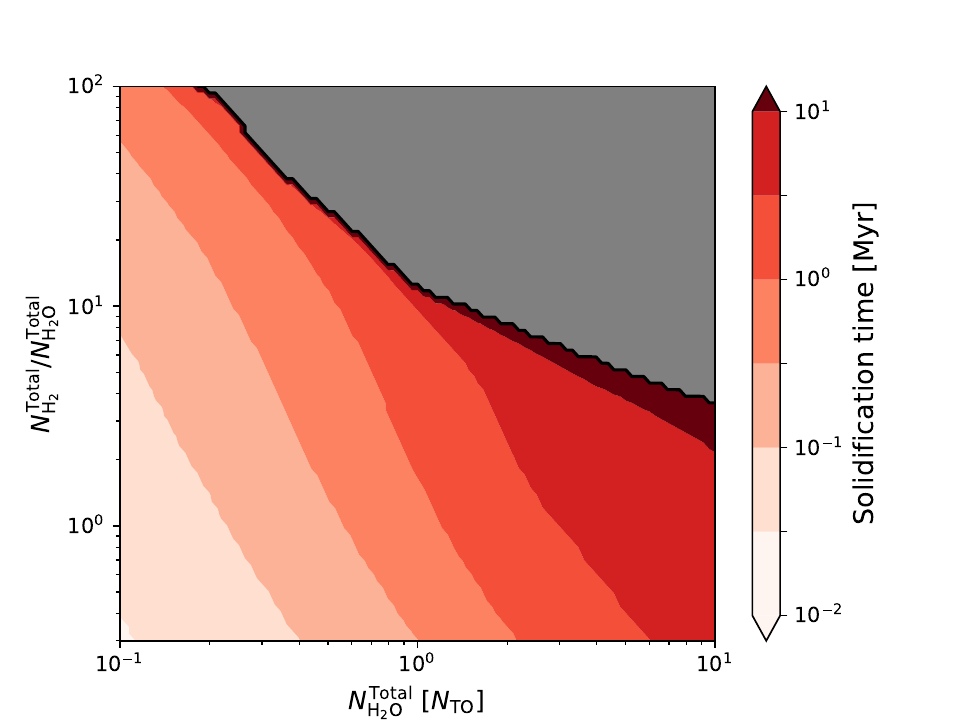}{0.45\textwidth}{(a)}
		\fig{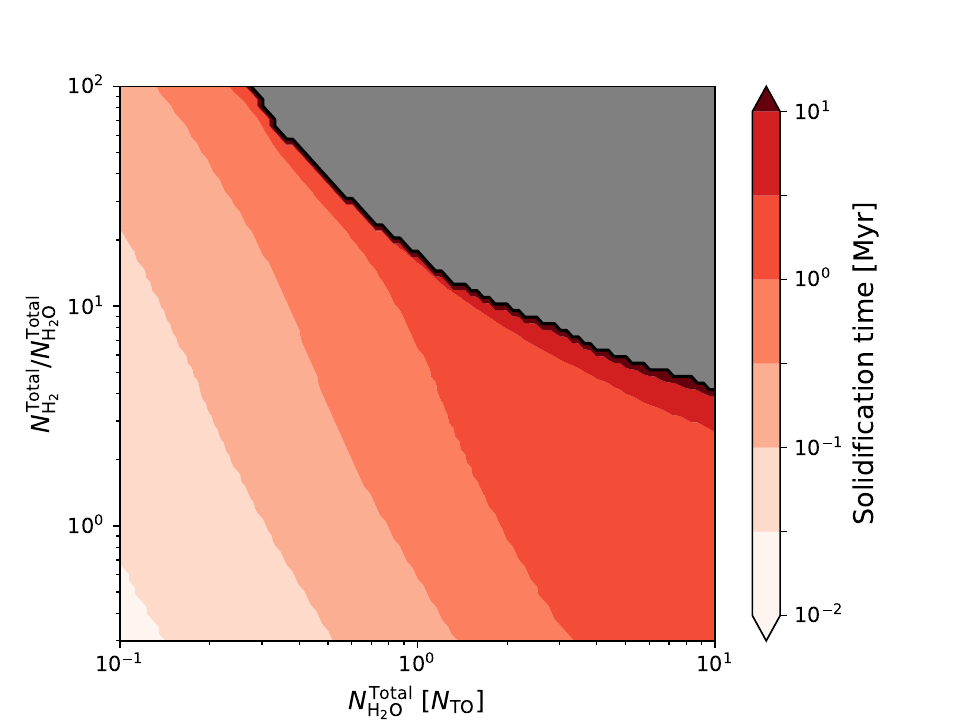}{0.45\textwidth}{(b)}
		}		
\caption{Time required for the temperature at a pressure of 23 GPa to reach the solidus temperature as a function of the total amounts of H$_{2}$O and H$_{2}$. (a) and (b) are the results with and without the scattering blanketing effects, respectively. The horizontal axis is the total amount of H$_2$O normalized by the molecular number of hydrogen in the present-day terrestrial seawater ($N_{\rm TO}$). The vertical axis is the ratio of the total amount of H$_2$ to that of H$_2$O. In the grey region, the equilibrium temperature at this region exceeds the solidus temperature.}
\label{fig:10}
\end{figure}

	The prolonged lifetime of the magma ocean due to the blanketing effect is expected to influence its chemical differentiation. \citet{Abe1997} theoretically demonstrated that differentiation between compatible and incompatible elements tends to be limited in the lower mantle compared to the upper mantle due to the rapid cooling not enough for melt-solid separation to take place in the case without a proto-atmosphere. On the other hand, the effective blanketing effect of a proto-atmosphere can prolong the magma ocean lifetime even in the high-pressure regions of the lower mantle.  Figure~\ref{fig:10} shows the time required for the temperature at a pressure of 23 GPa, which is the boundary between the lower and upper mantle, to reach the solidus temperature. The solidification time increases with the abundances of H$_2$O and H$_2$, reaching or exceeding $\sim 1\,\mathrm{Myr}$ which is comparable with the time required for melt-solid separation to take place in the upper mantle estimated by \citet{Abe1997}, across a wide range of initial conditions. These results indicate the differentiation between compatible and incompatible elements can efficiently occur in the lower mantle region as well as the upper mantle under a thick proto-atmosphere. In the lower mantle region, bridgmanite is expected to be the first phase to crystallize from the magma ocean \citep[e.g.,][]{Boukare2015, Caracas2019}. Efficient crystal settling could result in a chemical composition distinct from the upper mantle, such as a different Mg/Si ratio, as indicated by some geochemical studies of igneous rocks derived from plumes \citep[e.g.,][]{Murakami2024}. Additionally, differentiation in the lower mantle may be related to chemical heterogeneities reflected in variations in seismic velocities \citep[e.g.,][]{Garnero2008}.

\subsection{Implications of the prolonged magma ocean lifetime for Moon formation}
	The extension of the magma ocean phase due to the blanketing effect could have crucial implications for the Moon's formation. Precise geochemical measurements of lunar samples reveal a remarkable similarity in isotopic compositions between the Moon and Earth \citep[e.g.,][]{Melosh2014}. However, numerical models of the Moon-forming giant impact suggest that most of the Moon's material originated from the impactor \citep[e.g.,][]{Canup2004b}, making it challenging to naturally explain this isotopic resemblance to Earth. To address this discrepancy, \citet{Karato2014} and \citet{Hosono2019} proposed that a giant solid impactor hit the proto-Earth while it was covered with a magma ocean. In such a scenario, a substantial portion of the ejected Moon-forming material would derive from this magma ocean, aligning with the observed isotopic similarity between the Moon and Earth. Additionally, this model can account for the Moon’s higher FeO content compared to bulk silicate Earth, as FeO preferentially partitions into melt over coexisting minerals \citep[e.g.,][]{Mibe2006}. 
	
\begin{figure}
\gridline{\fig{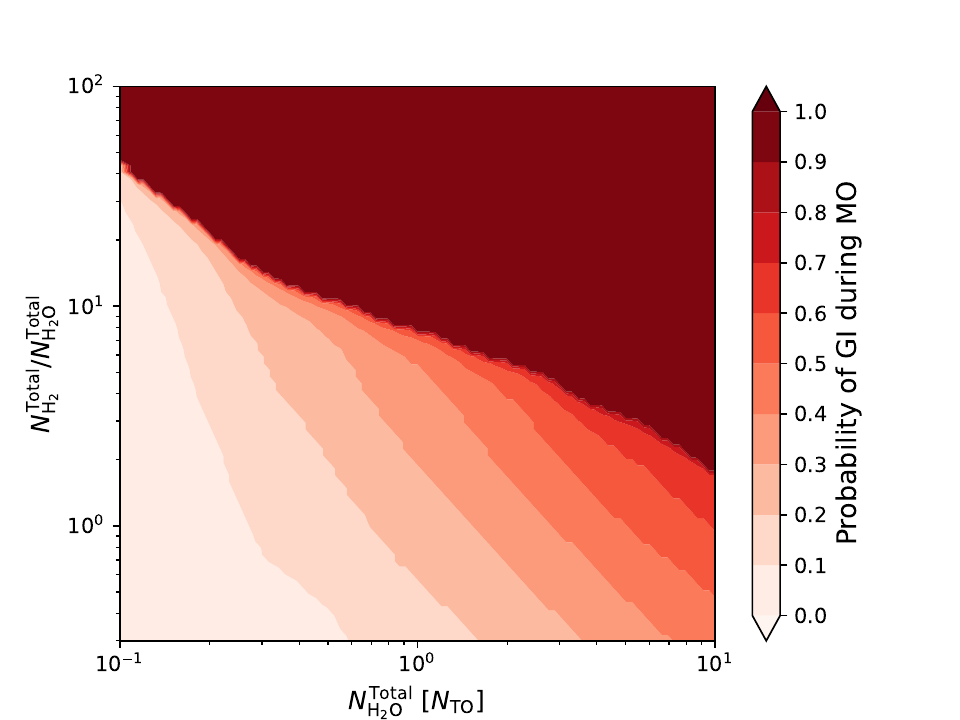}{0.45\textwidth}{(a)}
		\fig{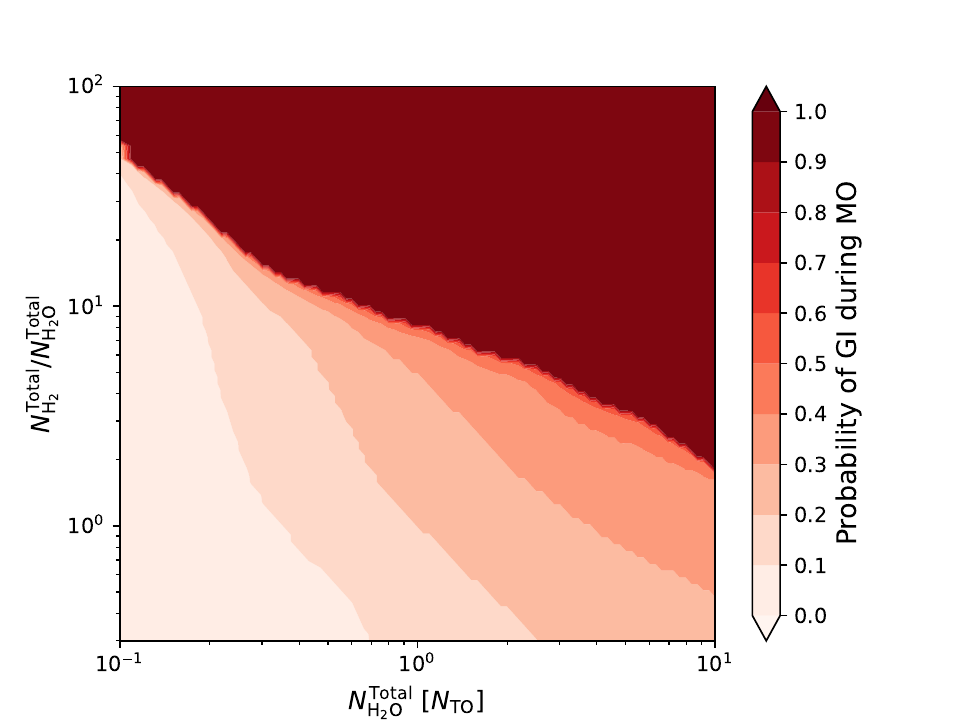}{0.45\textwidth}{(b)}
		}		
\caption{Probability of at least one giant impact occurring during the magma ocean phase, as described by Equation (26). The average time interval between giant impacts is set at 10 Myr. (a) and (b) are the results with and without the scattering blanketing effects, respectively. The horizontal axis is the total amount of H$_2$O normalized by the molecular number of hydrogen in the present-day terrestrial seawater ($N_{\rm TO}$). The vertical axis is the ratio of the total amount of H$_2$ to that of H$_2$O.}
\label{fig:11}
\end{figure}

	To assess the influence of the proto-atmosphere's blanketing effect, we estimate the probability of giant impacts occurring during the magma ocean phase. Geochronological data on lunar samples such as Hf--W isotopic systematics indicates that the Moon formed by a giant impact about 50--150 Myr after the formation of CAIs \citep[e.g.,][]{Jacobson2005, Jacobson2014, Fischer2018, Thiemens2019, Canup2023}. Supposing that Earth experienced about ten times giant impacts, the average time interval of giant impacts is expected to have been 5--15 Myr. Figure~\ref{fig:11} shows the probability $P_{\rm imp}$ that one or more giant impacts occurred during the magma ocean phase, calculated using a Poisson distribution:
	\begin{equation}
		P_{\rm imp}=1-\mathrm{exp}\left(-\frac{\tau_{\rm MO}}{\tau_{\rm imp}}\right),
	\end{equation}
	where $\tau_{\rm MO}$ is the magma ocean lifetime in Figure~\ref{fig:6} and $\tau_{\rm imp}$ is the average interval between giant impacts. Here $\tau_{\rm imp}$ is set to be 10 Myr, and $\tau_{\rm MO}$ is assumed to be 50 Myr under equilibrium conditions between OLR and net solar absorption (grey regions in Figure~\ref{fig:6}). The prolonged magma ocean lifetime due to the scattering blanketing effect increases the probability of a giant impact occurring during the magma ocean phase. For instance, the probability increases from 0.34 to 0.62 by the scattering blanketing effect when the total amounts of both H$_2$O and H$_2$ are $10N_{\rm TO}$. This enhanced probability provides a compelling explanation for the Moon's observed chemical and isotopic characteristics, supporting the hypothesis of a magma ocean origin of the Moon.
	
\section{Conclusions}
	We developed a 1-D radiative transfer model for planetary and solar radiation in proto-atmospheres composed of H$_2$O and H$_2$ and a coupled thermal evolution model of interiors and proto-atmospheres to investigate the scattering blanketing effect on planetary radiation and magma ocean cooling. Our results show that Rayleigh scattering significantly reduces outgoing planetary radiation at wavelengths below $\sim 1\,\mathrm{\micron}$, particularly in hot, thick atmospheres where scattering is highly effective. Consequently, the planetary outgoing radiation flux decreases by up to about one to two orders of magnitude, and the magma ocean lifetime is prolonged by up to about three times due to the scattering blanketing effect when the total amounts of H$_2$O and H$_2$ are equivalent to or greater than the present-day terrestrial seawater. These findings suggest that the prolonged magma ocean phase facilitated efficient differentiation between compatible and incompatible elements, even in the lower mantle. Furthermore, they imply that a sustained magma ocean likely persisted throughout much of the giant impact phase, supporting a magma ocean origin of the Moon consistent with its observed chemical characteristics.
	
\bibliographystyle{aasjournal} % style aa.bst
\bibliography{references} 	
	
%\appendix
%\section{Details of the model}



\end{document}